\documentclass[11pt,journal,onecolumn,draftcls]{IEEEtran}

\IEEEoverridecommandlockouts

\usepackage{hyperref}
\usepackage{xr}
\usepackage[table]{xcolor}
\usepackage{array}
\usepackage{bbding}
\usepackage{cite}
\usepackage{graphicx}
\usepackage{amssymb}
\usepackage{amsmath}
\usepackage{amsfonts}
\usepackage{algorithmic}
\usepackage{algorithm}
\usepackage{mdwmath}
\usepackage{mdwtab}
\usepackage{eqparbox}
\usepackage{fixltx2e}
\usepackage{stfloats}
\usepackage{dsfont}
\usepackage{mathrsfs}
\usepackage{bbm}
\usepackage{bm}
\usepackage{threeparttable}
\usepackage{multirow}
\usepackage[caption=false,font=footnotesize]{subfig}
\usepackage{centernot}
\usepackage{accents}
\usepackage{cleveref}
\usepackage{lscape}

\newtheorem{theorem}{Theorem}
\newtheorem{assumption}{Assumption}
\newtheorem{lemma}{Lemma}

\newcommand{\qed}{\nobreak \ifvmode \Relax \else
      \ifdim\lastskip<1.5em \hskip-\lastskip
      \hskip1.5em plus0em minus0.5em \fi \nobreak
      \vrule height0.65em width0.65em depth0em\fi}

\def\defeq{\triangleq}

\DeclareMathOperator*{\minimize}{\mathrm{minimize}}

\def\ds{\mathds}
\def\wt{\widetilde}
\def\wh{\widehat}

\def\col{{\mathrm{col}}}
\def\row{{\mathrm{row}}}
\def\diag{{\mathrm{diag}}}

\def\Tr{{\mathrm{Tr}}}

\def\kron{\otimes}

\def\glob{\textrm{glob}}

\def\E{\mathbb{E}}
\def\F{\mathbb{F}}

\def\nn{{\nonumber}}

\newcommand{\be}{\begin{equation}}
\newcommand{\ee}{\end{equation}}
\newcommand{\bea}{\begin{eqnarray*}}
\newcommand{\eea}{\end{eqnarray*}}

\makeatletter
\def\Lddots{\mathinner{\mkern1mu\raise17\p@\vbox{\kern17\p@\hbox{.}}\mkern2mu
    \raise8\p@\hbox{.}\mkern2mu\raise\p@\hbox{.}\mkern1mu}}
 \makeatother

\newcommand{\mbbB}{\mathbb{B}}
\newcommand{\mbbC}{\mathbb{C}}

\newcommand{\mbbR}{\mathbb{R}}
\newcommand{\mbbT}{\mathbb{T}}

\newcommand{\Acal}{\mathcal{A}}

\newcommand{\Hcal}{\mathcal{H}}

\newcommand{\Mcal}{\mathcal{M}}
\newcommand{\Ncal}{\mathcal{N}}

\newcommand{\Rcal}{\mathcal{R}}

\def\Re{\mathfrak{Re}}
\def\Im{\mathfrak{Im}}

\def\Nknk{\Ncal_k\backslash\{k\}}

\newcommand{\one}{\ds{1}}

\newcommand{\T}{\mathsf{T}} 
\newcommand{\ubar}{\underaccent{\bar}}

\begin{document}
%
\title{Asynchronous Adaptation and Learning over Networks --- Part I: Modeling and Stability Analysis}

\author{Xiaochuan~Zhao,~\IEEEmembership{Student~Member,~IEEE,}
        and Ali~H.~Sayed,~\IEEEmembership{Fellow,~IEEE} \\
\thanks{The authors are with Department of Electrical Engineering, University of California, Los Angeles, CA 90095
Email: \{xzhao,~sayed\}@ee.ucla.edu.}
\thanks{This work was supported by NSF grants CCF-1011918 and ECCS-1407712. A short and limited early version of this work appeared in the conference proceeding \cite{Zhao12EUSIPCO}. Parts II and III of this work are presented in \cite{Zhao13TSPasync2, Zhao13TSPasync3}.}}


\maketitle

\begin{abstract}
In this work and the supporting Parts II \cite{Zhao13TSPasync2} and III \cite{Zhao13TSPasync3}, we provide a rather detailed analysis of the stability and performance of asynchronous strategies for solving distributed optimization and adaptation problems over networks. We examine asynchronous networks that are subject to fairly general sources of uncertainties, such as changing topologies, random link failures, random data arrival times, and agents turning on and off randomly. Under this model, agents in the network may stop updating their solutions or may stop sending or receiving information in a random manner and without coordination with other agents. We establish in Part I conditions on the first and second-order moments of the relevant parameter distributions to ensure mean-square stable behavior. We derive in Part II \cite{Zhao13TSPasync2} expressions that reveal how the various parameters of the asynchronous behavior influence network performance. We compare in Part III \cite{Zhao13TSPasync3} the performance of asynchronous networks to the performance of both centralized solutions and synchronous networks. One notable conclusion is that the mean-square-error performance of asynchronous networks shows a degradation only of the order of $O(\nu)$, where $\nu$ is a small step-size parameter, while the convergence rate remains largely unaltered. The results provide a solid justification for the remarkable resilience of cooperative networks in the face of random failures at multiple levels: agents, links, data arrivals, and topology.
\end{abstract}


\allowdisplaybreaks

\section{Introduction}
Distributed learning arises when a global objective needs to be achieved through local cooperation over a number of interconnected agents. This problem occurs in many important contexts, including distributed estimation \cite{Dimakis10PROC, Kar11TSP, Bertrand12TSP, Gharehshiran13JSTSP, Lopes08TSP, Cattivelli10TSP, Xiao06IPSN, Sayed14PROC}, distributed machine learning \cite{Predd09TIT, Theodoridis11SPM, Chouvardas11TSP, Chen11TSP}, resource allocation \cite{Gesbert07Proc, Lorenzo11TSP}, and in the modeling of flocking and swarming behavior by biological networks \cite{Passino02CSM, Olfati06TAC, Barbarossa07SPM, Cattivelli11TSP, Tu11JSTSP}. Several useful decentralized solutions, such as consensus strategies \cite{Tsitsiklis84TAC, Tsitsiklis86TAC, Xiao04SCL, Boyd06TIT, Braca08FUSION, Nedic09TAC, Kar09TSP, Srivastava11JSTSP, Hlinka12TSP}, incremental strategies \cite{Bertsekas97JOP, Nedic01JOP, Rabbat05JSAC, Blatt07SIAM, Lopes07TSP}, and diffusion strategies \cite{Lopes08TSP, Cattivelli10TSP, Chen11TSP, Sayed13SPM, Sayed13Chapter, Sayed14PROC, Sayed14NOW}, have been developed for this purpose. Diffusion strategies are particularly attractive because they are scalable, robust, fully-distributed, and endow networks with real-time adaptation and learning abilities. In addition, they have been shown to have superior stability ranges \cite{Tu12TSP} and to lead to enhanced transient and steady-state performance in the context of \emph{adaptive} networks when \emph{constant} step-sizes are necessary to enable continuous adaptation and learning. The main reason for this enhanced performance can be explained as follows. In adaptive implementations, the individual agents do not know their exact cost functions and need to estimate their respective gradient vectors. The difference between the actual and approximate gradient vectors is called gradient noise. When a constant step-size is used in a stochastic-gradient implementation, the gradient noise term does not die out anymore and seeps into the algorithm. This effect disappears when a diminishing step-size is used because the decaying step-size annihilates the gradient noise factor. However, it was shown in \cite{Tu12TSP} that when constant step-sizes are used to enable adaptation, the effect of gradient noise can make the state of consensus networks grow unstable. The same effect does not happen for diffusion networks; the stability of these networks was shown to be insensitive to the network topology. This is an important property, especially for asynchronous networks where the topology will be changing randomly. It therefore becomes critical to rely on distributed strategies that are robust to such changes. For this reason, we concentrate on the study of diffusion networks while noting that most of the analysis can be extended to consensus networks with some minimal adjustments. In this Part I, we show that diffusion strategies continue to deliver stable network behavior under fairly general asynchronous conditions for non-vanishing step-size adaptation.

There already exist several insightful studies and results in the literature on the performance of consensus and gossip type strategies in the presence of asynchronous events \cite{Tsitsiklis86TAC, Boyd06TIT, Kar09TSP, Srivastava11JSTSP} or changing topologies \cite{Boyd06TIT, Kar08TSP, Kar09TSP, Aysal09Allerton, Aysal09TSP, Kar10TSP, Jakovetic10TSP, Jakovetic11TSP, Kar11TSP, Srivastava11JSTSP}. There are also some limited studies in the context of diffusion strategies \cite{Lopes08ICASSP, Takahashi10ICASSP}. However, with the exception of the latter two works, the earlier references investigated pure averaging algorithms \emph{without} the ability to respond to streaming data, assumed noise-free data, or relied on the use of diminishing step-size sequences. These conditions are problematic for adaptation and learning purposes when data is continually streaming in, since decaying step-sizes turn off adaptation eventually, and noise (including gradient noise) is always present.

In this article, and its accompanying Parts II \cite{Zhao13TSPasync2} and III \cite{Zhao13TSPasync3}, we remove these limitations. We also allow for fairly general sources of uncertainties and random failures and permit them to occur simultaneously. Some of the questions that we address in the three parts include:
\begin{enumerate}
\item How does asynchronous behavior affect network stability? Can mean-square stability still be ensured under non-vanishing step-sizes?

\item How is the convergence rate of the algorithm affected? Is it altered relative to synchronous networks?

\item Are agents still able to reach some sort of agreement in steady-state despite the random nature of their interactions and despite data arriving at possibly different rates?

\item How close do the steady-state iterates of the various agents get to each other and to the optimal solution that the network is seeking?

\item Compared with synchronous networks, under what conditions and by how much does the asynchronous behavior generate a \emph{net} negative effect in performance?

\item How close can the performance of an asynchronous network get to that of a stochastic-gradient centralized solution?
\end{enumerate}

\noindent We answer question 1 in Part I, questions 2, 3, and 4 in Part II \cite{Zhao13TSPasync2}, and questions 5 and 6 in Part III \cite{Zhao13TSPasync3}. As the reader can ascertain from the derivations in the appendices, the arguments require some careful analysis due to various sources of randomness in the network and due to the interaction among the agents --- events at one agent influence the operation of other neighboring agents.

To answer the above questions in a systematic manner, we introduce in this Part I a fairly general model for asynchronous events. Then,  we carry out a detailed mean-square-error (MSE) analysis and arrive at explicit conditions on the parameters of certain probability distributions to ensure stable behavior. The analysis is pursued further to arrive at closed-form expressions for the MSE in steady-state in Part II \cite{Zhao13TSPasync2} and to compare against synchronous and centralized environments in Part III \cite{Zhao13TSPasync3}. One of the main conclusions that will follow from the detailed analysis in the three parts is that, under certain reasonable conditions, the asynchronous network will continue to be able to deliver performance that is comparable to the synchronous case where no failure occurs. This work therefore justifies analytically why, even under highly dynamic scenarios where many different components of the network can vary randomly or fail, the diffusion network is still able to deliver performance and solve the inference or optimization task with remarkable accuracy. The results provide strong evidence for the intrinsic robustness and resilience of network-based cooperative solutions. When presenting the material in the three parts, we focus on discussing the main results and their interpretation in the body of the paper, while delaying the technical proofs and arguments to the appendices.

\emph{Notation}: We use lowercase letters to denote vectors, uppercase letters for matrices, plain letters for deterministic variables, and boldface letters for random variables. We also use $(\cdot)^{\T}$ to denote transposition, $(\cdot)^{*}$ to denote conjugate transposition, $(\cdot)^{-1}$ for matrix inversion, $\Tr(\cdot)$ for the trace of a matrix, $\lambda(\cdot)$ for the eigenvalues of a matrix, and $\| \cdot \|$ for the 2-norm of a matrix or the Euclidean norm of a vector. Besides, we use $\kron$ to denote the Kronecker product.

\section{Preliminaries}
\label{sec:pre}
\begin{figure}
\includegraphics[scale=0.4]{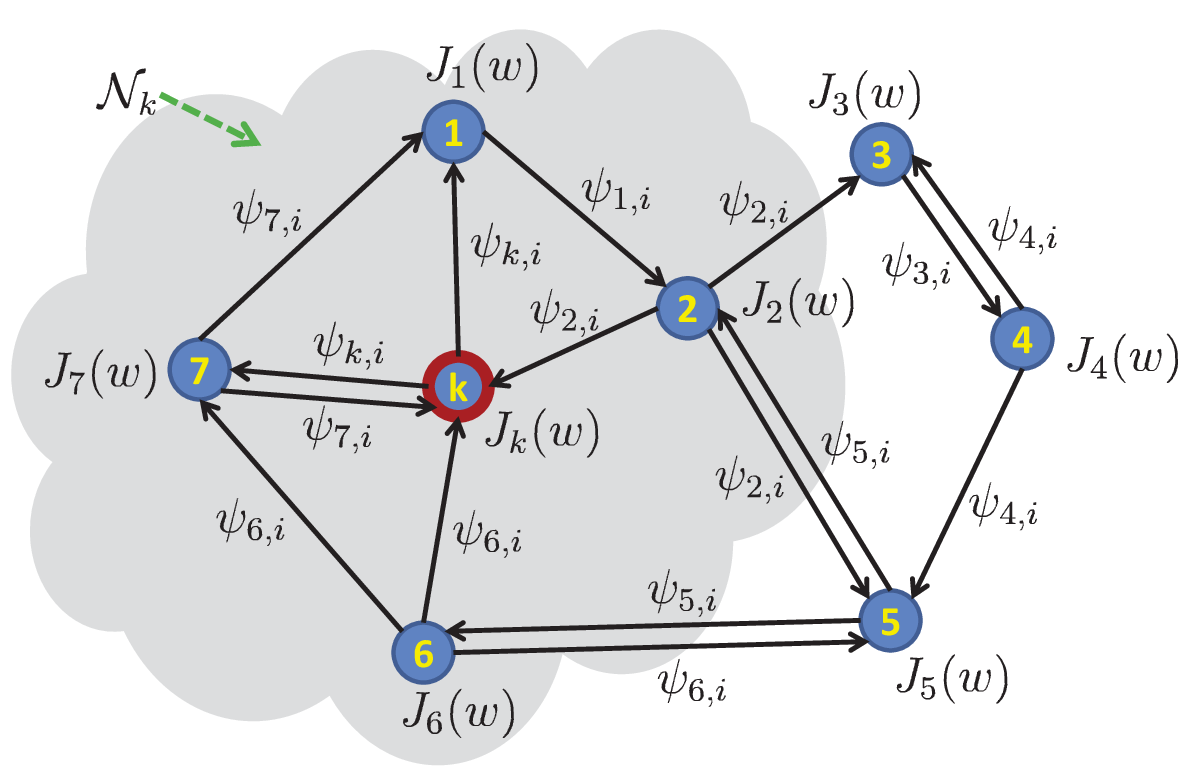}
\centering
\caption{An illustration of a connected network with individual costs associated with the various agents.}
\label{fig:network_illustration}
\vspace{-1\baselineskip}
\end{figure}

We consider a connected network consisting of $N$ agents as shown in Fig.~\ref{fig:network_illustration}. The objective is to minimize, in a distributed manner, an aggregate cost function of the form:
\begin{align}
\label{eqn:globalcost}
\minimize_{w}\;\;J^{\glob}(w)\defeq\sum_{k=1}^{N}J_k(w)
\end{align}
where the $\{J_k(w) : w\in\mbbC^{M} \rightarrow \mbbR\}$ denote individual cost functions. Observe that we are allowing the argument $w$ to be complex-valued so that the results are applicable to a wide range of problems, especially in the fields of communications and signal processing where complex parameters are fairly common (e.g., in modeling wireless channels, power grid models, beamforming weights, etc.). To facilitate the analysis, and before describing the distributed strategies, we first introduce two alternative ways for representing real-valued functions of complex arguments.

\subsection{Equivalent Representations}
\label{subsec:equivalent}
The first representation is based on the \emph{1-to-1} mapping $\bar\mbbT: \mbbC^M \longmapsto \mbbR^{2M}$:
\be
\label{eqn:bardef}
\bar{w} \defeq \bar\mbbT(w) = \begin{bmatrix}
\Re(w) \\
\Im(w)
\end{bmatrix}
\ee
which replaces the $M \times 1$ complex vector $w$ by the $2M \times 1$ extended vector 
$\bar{w}$ composed of the real and imaginary components of $w$. In this way, we can interpret each $J_k(w)$ as a function of the real-valued variable $\bar{w}$ and write $J_k(\bar{w}) \defeq J_k(w)$ as well as
\be
\label{eqn:globalcostreal}
J^\glob(\bar w) \defeq J^\glob(w) = \sum_{k=1}^{N}J_k(w) = \sum_{k=1}^{N} J_k(\bar w)
\ee
The second representation for functions of complex arguments is based on another \emph{1-to-1} mapping $\ubar\mbbT: \mbbC^M \longmapsto \mbbC_M^{2M}$ (where $\mbbC_M^{2M}$ is a sub-manifold of complex dimension $M$ and is isomorphic to $\mbbR^{2M}$ \cite{Delgado09}) defined as
\be
\label{eqn:ubardef}
\ubar{w} \defeq \ubar\mbbT(w) = \begin{bmatrix}
w \\
(w^*)^{\T}
\end{bmatrix}
\ee
in terms of the entries of $w$ and their complex conjugates. In this case, we can interpret each $J_k(w)$ as a function defined over the extended variable $\ubar{w}\in\mbbC_M^{2M}$ and write $J_k(\ubar w) \defeq J_k(w)$ as well as
\be
\label{eqn:globalcostcomp}
J^\glob(\ubar w) \defeq J^\glob(w) = \sum_{k=1}^{N}J_k(w) = \sum_{k=1}^{N} J_k(\ubar w)
\ee
Most of our analysis will be based on the second representation \eqref{eqn:ubardef}--\eqref{eqn:globalcostcomp}; the first representation \eqref{eqn:bardef}--\eqref{eqn:globalcostreal} will be used when we need to exploit some analytic properties of real functions. Note from \eqref{eqn:bardef} and \eqref{eqn:ubardef} that the variables $\{\bar{w},\ubar{w}\}$ are related linearly as follows:
\be
\label{eqn:barw2ubarw}
\underbrace{\begin{bmatrix}
w \\
(w^*)^{\T}
\end{bmatrix}}_{ = \ubar{w} } = \underbrace{\begin{bmatrix}
I_M \,&\, jI_M \\
I_M \,&\, -jI_M
\end{bmatrix}}_{\defeq D} \underbrace{\begin{bmatrix}
\Re(w) \\
\Im(w)
\end{bmatrix}}_{ = \bar{w}} \Longleftrightarrow \ubar w = D \cdot \bar w
\ee
where the matrix $D$ satisfies $DD^* = D^* D = 2 \cdot I_{2M}$ and $I_{2M}$ denotes the $2M\times 2M$ identity matrix. It follows that
\be
\label{eqn:wbartowubar}
\bar{w} = D^{-1} \cdot \ubar{w} = \frac{1}{2} D^* \cdot \ubar{w}
\ee
Using the real representation $\{J_k(\bar{w})\}$, we introduce the following assumption on the analytic properties of $\{J_k(w)\}$.

\begin{assumption}[{Properties of cost functions}]
\label{asm:costfunctions}
The individual cost functions $\{J_k(\bar{w}):\mbbR^{2M} \mapsto \mbbR \,; k = 1, 2, \dots, N\}$ are assumed to be at least twice-differentiable and strongly convex over $\mbbR^{2M}$. They are also assumed to share a \emph{common} and \emph{unique} minimizer at $\bar{w}^o \defeq \bar\mbbT(w^o)$, where $w^o \in \mbbC^{M}$. \hfill \IEEEQED
\end{assumption}

The situation involving a common minimizer for the cost functions $\{J_k(\bar{w})\}$ is frequent in practice, especially when agents need to cooperate with each other in order to attain a \emph{common} objective. For example, in biological networks, it is usual for agents in a school of fish to interact while searching for a common food source or avoiding a common predator \cite{Tu11JSTSP}. Likewise, in wireless sensor networks, it is common for sensors to survey the same physical environment, to interact with each other to estimate a common modeling parameter, or to track the same target \cite{Lorenzo11TSP}. Furthermore, in machine learning applications \cite{Towfic13NC}, it is common for all agents to minimize the same cost function (for example, the logistic risk) which automatically satisfies the condition of a common minimizer. It is also important to note that agents can still benefit from cooperation even when they share a common minimizer for at least two reasons. First, this is because different agents are generally subject to different measurement noise conditions. The information sharing and cooperation among agents can effectively equalize the difference. Second, this is also because some agents may not have sufficient data to recover the desired unknown parameter on their own. Information sharing and cooperation among agents can alleviate the problem of ill-conditioning and enable agents to solve for their desired parameters.

It follows from Assumption \ref{asm:costfunctions} that the real global cost function, $J^\glob(\bar{w})$, has a unique minimizer at $\bar w^o$, or equivalently, that the original global cost function, $J^\glob(w)$, has a unique minimizer at $w^o$. Accordingly, the unique minimizer for $J^\glob(\ubar w)$ and for each $J_k(\ubar w)$ is given by $\ubar w^o = \ubar\mbbT(w^o)$.

The strong convexity assumption on each cost $J_k(\bar{w})$ ensures that their Hessian matrices are sufficiently bounded away from zero, which avoids situations involving ill-conditioning in recursive implementations based on streaming data. Strong convexity is not a serious limitation because it is common practice in adaptation and learning to incorporate regularization into the cost functions, and it is well-known that regularization helps enforce strong convexity \cite{Boyd04, Sayed08}. We may add though that many of the results in this work would still hold if we only require the aggregate cost function $J^\glob(\bar{w})$ to be strongly convex by following arguments similar to those used in \cite{Chen13JSTSP}; in that case, it would be sufficient to require only one of the individual costs $J_k(\bar{w})$ to be strongly convex while the remaining costs can be simply convex. Nevertheless, some of the derivations will become more technical under these more relaxed conditions. For this reason, and since the arguments in the three parts are already demanding and lengthy, we opt to convey the main ideas and results by working under Assumption \ref{asm:costfunctions}.

\subsection{Hessian Matrices}

We explain in Appendix \ref{app:complex} how to compute the complex gradient vector and the complex Hessian matrix of the cost $J_k(w)$, and its equivalent representations, with respect to their arguments. The strong convexity condition from Assumption \ref{asm:costfunctions} translates into the existence of a lower bound on the Hessian matrices as shown below in \eqref{eqn:boundseigHessian}. In addition, we shall assume that the Hessian matrices are also bounded from above. This requirement relaxes conditions from prior studies in the literature where it has been customary to bound the gradient vector as \emph{opposed} to the Hessian matrix \cite{Nedic09TAC, Srivastava11JSTSP}; bounding the gradient vector limits the class of cost functions to those with linear growth --- see \cite{Chen11TSP, Sayed14NOW} for an explanation. 

\begin{assumption}[{Bounded Hessian and Lipschitz condition}]
\label{asm:boundedHessian}
The eigenvalues of the complex Hessian $\{\nabla_{\ubar{w}\ubar{w}^*}^2 J_k(\ubar{w})\}$ (defined by \eqref{eqn:complexHessianextdef} in Appendix \ref{app:complex}) are bounded from below and from above by
\be
\label{eqn:boundseigHessian}
\lambda_{k,\min} \le \lambda( \nabla_{\ubar{w}\ubar{w}^{*}}^2 J_k(\ubar{w})) \le \lambda_{k,\max}
\ee
where $0 < \lambda_{k,\min} \le \lambda_{k,\max}$. Moreover, the complex Hessian functions $\{\nabla_{\ubar{w}\ubar{w}^{*}}^2 J_k(\ubar{w}\}$ are assumed to be \emph{locally} Lipschitz continuous \cite{Polyak87} at $\ubar{w}^o$, i.e.,
\be
\label{eqn:Lipschitz}
\| \nabla_{\ubar{w}\ubar{w}^{*}}^2 J_k(\ubar{w}^o) - \nabla_{\ubar{w}\ubar{w}^{*}}^2 J_k(\ubar{w}) \| \le \tau_k \cdot \| \ubar{w}^o - \ubar{w} \|
\ee
where $\tau_k  \ge  0$, $\ubar{w}^o  =  \ubar{\mbbT}(w^o)$ and $\ubar{w}  =  \ubar{\mbbT}(w)$ for any $w \in {\mbbB}({w}^o, \delta_k)$ with ${\mbbB}({w}^o, \delta_k)$ denoting the 2-norm ball ${\mbbB}({w}^o, \delta_k) \defeq \{w \in \mbbC^{M}; \|w^o - w\| \le \delta_k\}$, which is centered at $w^o$ with radius $\delta_k$. \hfill \IEEEQED
\end{assumption}

\begin{lemma}[Global Lipschitz continuity]
\label{lemma:lipschitzHessian}
When conditions \eqref{eqn:boundseigHessian} and \eqref{eqn:Lipschitz} hold, the Hessian matrix functions $\{\nabla_{\ubar{w}\ubar{w}^{*}}^2 J_k(\ubar{w})\}$ are globally Lipschitz continuous at $w^o$, i.e.,
\be
\label{eqn:globalLipschitz}
\| \nabla_{\ubar{w}\ubar{w}^{*}}^2 J_k(\ubar{w}^o) - \nabla_{\ubar{w}\ubar{w}^{*}}^2 J_k(\ubar{w}) \| \le \tau_k' \cdot \| \ubar{w}^o - \ubar{w} \|
\ee
for any $w$, and 
\be
\tau_k' \defeq \max \left\{ \tau_k, \frac{\lambda_{k,\max} - \lambda_{k,\min}}{\sqrt{2}\delta_k} \right\}
\ee
\end{lemma}
\begin{IEEEproof}
We first note that 
\be
\label{eqn:boundedHessian1}
\| \nabla_{\ubar{w}\ubar{w}^{*}}^2 J_k(\ubar{w}^o) - \nabla_{\ubar{w}\ubar{w}^{*}}^2 J_k(\ubar{w}) \| \le \lambda_{k,\max} - \lambda_{k,\min}
\ee
for any $\ubar{w} = \ubar{\mbbT}(w)$, because for any $2M\times1$ vector $x$, 
\begin{align}
x^*[\nabla_{\ubar{w}\ubar{w}^{*}}^2 J_k(\ubar{w}^o) - \nabla_{\ubar{w}\ubar{w}^{*}}^2 J_k(\ubar{w})]x 
& = x^*[\nabla_{\ubar{w}\ubar{w}^{*}}^2 J_k(\ubar{w}^o)]x  - x^*[\nabla_{\ubar{w}\ubar{w}^{*}}^2 J_k(\ubar{w})]x \nn \\
& \le (\lambda_{k,\max} - \lambda_{k,\min}) \| x \|^2
\end{align}
Now, if $w\in {\mbbB}({w}^o, \delta_k)$, by condition \eqref{eqn:Lipschitz}, we have
\be
\| \nabla_{\ubar{w}\ubar{w}^{*}}^2 J_k(\ubar{w}^o) - \nabla_{\ubar{w}\ubar{w}^{*}}^2 J_k(\ubar{w}) \|  \le \tau_k' \cdot \| \ubar{w}^o - \ubar{w} \|
\ee
On the other hand, if $w \notin {\mbbB}({w}^o, \delta_k)$, i.e., $\| w^o - w \| > \delta_k$ or $\| \ubar{w}^o - \ubar{w} \| > \sqrt{2} \delta_k$, then we have
\begin{align}
\| \nabla_{\ubar{w}\ubar{w}^{*}}^2 J_k(\ubar{w}^o) - \nabla_{\ubar{w}\ubar{w}^{*}}^2 J_k(\ubar{w}) \| & \le \frac{\lambda_{k,\max} - \lambda_{k,\min}}{\sqrt{2}\delta_k} \cdot \sqrt{2}\delta_k \nn \\
& \le \tau_k' \cdot \| \ubar{w}^o - \ubar{w} \|
\end{align}
by condition \eqref{eqn:boundedHessian1}.
\end{IEEEproof}

\section{Asynchronous Diffusion Networks}
We first describe the traditional synchronous diffusion network studied in \cite{Chen11TSP, Sayed14PROC}, then we introduce the asynchronous network and derive some useful properties.

\subsection{Synchronous Diffusion Networks}
\label{subsec:syncDiff}
References \cite{Chen11TSP, Sayed14PROC} deal with the optimization of aggregate real functions of the form $J^\glob(\bar{w})$. Starting from equations (12)--(14) from \cite{Chen11TSP} and using \eqref{eqn:barw2ubarw} we can derive the following diffusion strategy for solving the distributed optimization problem \eqref{eqn:globalcost} with \emph{constant} step-sizes:
\begin{subequations}
\begin{alignat}{2}
\label{eqn:atcsync1}
{\bm{\psi}}_{k,i} & = {\bm{w}}_{k,i-1} - \mu_k \wh{\nabla_{w^*}J_k}(\bm{w}_{k,i-1}) & \quad & \mbox{(adaptation)}\\
\label{eqn:atcsync2}
{\bm{w}}_{k,i} & = \sum_{\ell\in{\Ncal}_k}a_{\ell k}\,{\bm{\psi}}_{\ell,i} & \quad & \mbox{(combination)}
\end{alignat}
\end{subequations}
where \eqref{eqn:atcsync1} is a stochastic gradient approximation step for self-learning and \eqref{eqn:atcsync2} is a convex combination step for social-learning. The iterate ${\bm{w}}_{k,i}$ is the estimate for $w^o$ that is computed by agent $k$ at iteration $i$. The iterate ${\bm{\psi}}_{k,i}$ is an intermediate solution that results from the adaptation step and will be shared with the neighbors in the combination step. The factor $\mu_k$ is a positive step-size parameter and the combination coefficients $\{a_{\ell k}\}$ are nonnegative parameters and are required to satisfy the following constraints:
\be
\label{eqn:topologyconstraints}
\sum_{\ell\in\Ncal_k}a_{\ell k} = 1, \;\;\mbox{and}\;\; 
\begin{cases}
a_{\ell k} > 0, & \mbox{if} \;\; \ell \in \Ncal_k \\
a_{\ell k} = 0, & \mbox{otherwise} \\
\end{cases}
\ee
where $\Ncal_k$ denotes the set of neighbors of agent $k$ including $k$ itself. If we collect these coefficients into an $N \times N$ matrix such that $[A]_{\ell k} = a_{\ell k}$, then condition \eqref{eqn:topologyconstraints} implies that $A$ is a left-stochastic matrix, written as $A^\T \one_N = \one_N$ where $\one_N$ is the $N\times1$ vector with all entries equal to one. 

In \eqref{eqn:atcsync1}, the stochastic approximation for the true gradient vector is used because, in general, agents do not have sufficient information to acquire the true gradients. The difference between the true and approximate gradients is called gradient noise, which is random in nature and seeps into the algorithm. That is why the variables $\{{\bm{w}}_{k,i}\}$ in \eqref{eqn:atcsync1}--\eqref{eqn:atcsync2} are random and are represented in boldface. We model the gradient noise, denoted by $\bm{v}_{k,i}(\bm{w}_{k,i-1})$, as an additive random perturbation to the true gradient vector, i.e.,
\be
\label{eqn:linearperturbationmodel}
\wh{\nabla_{w^*}J_k}(\bm{w}_{k,i-1}) = \nabla_{w^*}J_k(\bm{w}_{k,i-1}) + \bm{v}_{k,i}(\bm{w}_{k,i-1})
\ee
Let $\F_{i-1}$ denote the filtration to represent all information available up to iteration $i-1$. The conditional covariance of the individual gradient noise $\ubar{\bm{v}}_{k,i}(\bm{w}_{k,i-1})$ is given by
\begin{align}
\label{eqn:Rkidef}
R_{k,i}(\bm{w}_{k,i-1}) & \defeq \E [ \ubar{\bm{v}}_{k,i}(\bm{w}_{k,i-1}) \ubar{\bm{v}}_{k,i}^*(\bm{w}_{k,i-1}) | \F_{i-1} ] \nn \\
& = \begin{bmatrix}
R_{v,k,i}(\bm{w}_{k,i-1}) & R_{v,k,i}'(\bm{w}_{k,i-1})  \\
R_{v,k,i}'^{*}(\bm{w}_{k,i-1}) & R_{v,k,i}^\T(\bm{w}_{k,i-1})  \\
\end{bmatrix}
\end{align}
by using \eqref{eqn:ubardef}, where
\begin{align}
\label{eqn:vkicond_cov1}
R_{v,k,i}(\bm{w}_{k,i-1}) & \defeq \E [ \bm{v}_{k,i}(\bm{w}_{k,i-1}) \bm{v}_{k,i}^*(\bm{w}_{k,i-1}) | \F_{i-1} ] \\
\label{eqn:vkicond_cov2}
R_{v,k,i}'(\bm{w}_{k,i-1}) & \defeq \E [ \bm{v}_{k,i}(\bm{w}_{k,i-1}) \bm{v}_{k,i}^\T(\bm{w}_{k,i-1}) | \F_{i-1} ]
\end{align}
so that $R_{v,k,i}(\bm{w}_{k,i-1})$ is Hermitian positive semi-definite and $R_{v,k,i}'(\bm{w}_{k,i-1})$ is symmetric. Let further
\be
\label{eqn:bigvidef}
\ubar{\bm{v}}_i(\bm{w}_{i-1}) \defeq \col\{\ubar{\bm{v}}_{1,i}(\bm{w}_{1,i-1}),\dots,\ubar{\bm{v}}_{N,i}(\bm{w}_{N,i-1})\}
\ee
The conditional covariance of $\ubar{\bm{v}}_i(\bm{w}_{i-1})$ is denoted by
\be
\label{eqn:bigRidef}
\Rcal_i(\bm{w}_{i-1}) \defeq \E [ \ubar{\bm{v}}_i(\bm{w}_{i-1}) \ubar{\bm{v}}_i^*(\bm{w}_{i-1}) | \F_{i-1} ]
\ee

\begin{assumption}[Gradient noise model]
\label{asm:gradientnoise}
The gradient noise $\bm{v}_{k,i}(\bm{w}_{k,i-1})$, conditioned on $\F_{i-1}$, is assumed to be independent of any other random sources including topology, links, combination coefficients, and step-sizes. The conditional mean and variance of $\bm{v}_{k,i}(\bm{w}_{k,i-1})$ satisfy:
\begin{align}
\label{eqn:vkicond_mean}
\E\,[\bm{v}_{k,i}(\bm{w}_{k,i-1}) | \F_{i-1} ] & = 0 \\
\label{eqn:vkicond_var}
\E\,[\|\bm{v}_{k,i}(\bm{w}_{k,i-1})\|^2| \F_{i-1} ] & \le \alpha \, \|w^o - \bm{w}_{k,i-1}\|^2 + \sigma_v^2
\end{align}
for some $\alpha \ge 0$ and $\sigma_v^2 \ge 0$. \hfill \IEEEQED
\end{assumption}

Let $\ubar{\bm{v}}_{k,i}(\bm{w}_{k,i-1}) \defeq \ubar{\mbbT}(\bm{v}_{k,i}(\bm{w}_{k,i-1}))$. From Assumption \ref{asm:gradientnoise}, the extended gradient noise $\ubar{\bm{v}}_{k,i}(\bm{w}_{k,i-1})$, conditioned on $\F_{i-1}$, is independent of other random sources including topology, links, combination coefficients, and step-sizes. The conditional mean and variance of $\ubar{\bm{v}}_{k,i}(\bm{w}_{k,i-1})$ satisfy
\begin{align}
\label{eqn:gradientnoisemean}
\E[\ubar{\bm{v}}_{k,i}(\bm{w}_{k,i-1}) | \F_{i-1} ] & = 0 \\
\label{eqn:gradientnoise2norm}
\E[\|\ubar{\bm{v}}_{k,i}(\bm{w}_{k,i-1})\|^2| \F_{i-1} ] & \le \alpha \|\ubar{w}^o - \ubar{\bm{w}}_{k,i-1}\|^2 + 2\sigma_v^2
\end{align}
Conditions similar to \eqref{eqn:gradientnoisemean} and \eqref{eqn:gradientnoise2norm} appeared in the works \cite{Polyak87, Bertsekas99SIAM, Chen11TSP} on distributed algorithms. However, they are more relaxed than those employed in \cite{Polyak87, Bertsekas99SIAM}, as already explained in \cite{Chen11TSP}. Conditions \eqref{eqn:gradientnoisemean} and \eqref{eqn:gradientnoise2norm} are satisfied in several useful scenarios of practical relevance such as those involving quadratic costs or logistic costs.

\subsection{Asynchronous Diffusion Networks}
\label{sec:model}
To model the \emph{asynchronous} behavior of the network, we modify the diffusion strategy \eqref{eqn:atcsync1}--\eqref{eqn:atcsync2} to the following form:
\begin{subequations}
\begin{align}
\label{eqn:atcasync1}
{\bm{\psi}}_{k,i} & = {\bm{w}}_{k,i-1} - \bm{\mu}_k(i) \wh{\nabla_{w^*}J_k}(\bm{w}_{k,i-1}) \\
\label{eqn:atcasync2}
{\bm{w}}_{k,i} & = \sum_{\ell\in\bm{\Ncal}_{k,i}} \bm{a}_{\ell k}(i)\,{\bm{\psi}}_{\ell,i}
\end{align}
\end{subequations}
where the $\{\bm{\mu}_k(i),\bm{a}_{\ell\,k}(i)\}$ are now \emph{time-varying} and \emph{random} step-sizes and combination coefficients, and $\bm{\Ncal}_{k,i}$ denotes the \emph{random} neighborhood of agent $k$ at time $i$. The step-size parameters $\{\bm{\mu}_k(i)\}$ are nonnegative random variables, and the combination coefficients $\{\bm{a}_{\ell k}(i)\}$ are also nonnegative random variables, which are required to satisfy the following constraints (compare to \eqref{eqn:topologyconstraints}):
\be
\label{eqn:randomtopologyconstraints}
\sum_{\ell\in\bm{\Ncal}_{k,i}} \bm{a}_{\ell k}(i) = 1, \;\;\mbox{and}\;\;
\begin{cases}
\bm{a}_{\ell k}(i) > 0, & \mbox{if} \; \ell \in \bm{\Ncal}_{k,i} \\
\bm{a}_{\ell k}(i) = 0, & \mbox{otherwise} \\
\end{cases}
\ee
Let 
\begin{align}
\bm{w}_{i} & \defeq \col\{\bm{w}_{1,i}, \bm{w}_{2,i}, \dots, \bm{w}_{N,i}\} \\
\bm{\psi}_i & \defeq \col\{\bm{\psi}_{1,i}, \bm{\psi}_{2,i}, \dots, \bm{\psi}_{N,i}\} 
\end{align}
denote the collections of the iterates from across the network at time $i$. Let also 
\be
\bm{M}_i \defeq \diag\{\bm\mu_1(i),\bm\mu_2(i),\dots,\bm\mu_N(i)\} 
\ee
be the diagonal random step-size matrix at time $i$. We further collect the combination coefficients $\{\bm{a}_{\ell k}(i)\}$ at time $i$ into the matrix $\bm{A}_i\in\mbbR^{N\times N}$. The asynchronous network model consists of the following conditions on $\{{\bm{M}}_i,{\bm{A}}_i;i\ge0\}$:

\begin{enumerate}
\item  The stochastic process $\{\bm{M}_i,i\ge0\}$ consists of a sequence of diagonal random matrices with \emph{bounded} nonnegative entries, $\{\bm\mu_k(i) \in [0, \mu_{k}]\}$, where the upper bound $\mu_{k} > 0$ is a constant. The random matrix $\bm{M}_i$ is assumed to have constant mean $\bar{M}$ of size $N\times N$ and constant Kronecker-covariance matrix $C_M$ of size $N^2 \times N^2$, i.e.,
\begin{align}
\label{eqn:randstepsizemean}
\bar{M} & \defeq \E\,\bm{M}_i = \diag\{\bar{\mu}_1, \bar{\mu}_2, \dots, \bar{\mu}_N\} \\
\label{eqn:randstepsizemeanentry}
\bar{\mu}_k & \defeq \E\,\bm{\mu}_k(i) \\
\label{eqn:randstepsizecov}
C_M & \defeq \E\,[(\bm{M}_i-\bar{M})\kron(\bm{M}_i-\bar{M})] = \diag\{C_{\mu,1}, C_{\mu,2}, \dots, C_{\mu,N}\} \\
\label{eqn:randstepsizecovblock}
C_{\mu,k} & \defeq \E[(\bm{\mu}_k(i) - \bar{\mu}_k)(\bm{M}_i-\bar{M})] = \diag\{ c_{\mu,k,1}, c_{\mu,k,2}, \dots, c_{\mu,k,N} \}
\end{align}
where $\bar{\mu}_k$ denotes the $k$-th entry on the diagonal of $\bar{M}$, $C_{\mu,k}$ is a diagonal matrix and denotes the $k$-th block of size $N\times N$ on the diagonal of $C_M$, and $c_{\mu,k,\ell}$ denotes the $\ell$-th entry on the diagonal of $C_{\mu,k}$. The scalar $c_{\mu,k,\ell}$ represents the covariance between the step-sizes $\bm{\mu}_k(i)$ and $\bm{\mu}_\ell(i)$:
\be
\label{eqn:randstepsizecoventry}
c_{\mu,k,\ell} \defeq \E [(\bm{\mu}_k(i) - \bar{\mu}_k)(\bm{\mu}_\ell(i) - \bar{\mu}_\ell)]
\ee
When $\ell = k$, the scalar $c_{\mu,k,k}$ becomes the variance of $\bm{\mu}_k(i)$. Since the $\{\bar{\mu}_k\}$ are all finite positive numbers, the condition number of the matrix $\bar{M}$ is bounded by some finite positive constant $\kappa>0$, i.e.,
\be
\label{eqn:kappadef}
\frac{\max_k\{\bar{\mu}_k\}}{\min_k\{\bar{\mu}_k\}} \le \kappa
\ee

\item The stochastic process $\{\bm{A}_i,i\ge0\}$ consists of a sequence of left-stochastic random matrices, whose entries satisfy the conditions in \eqref{eqn:randomtopologyconstraints} at every time $i$. The mean and Kronecker-covariance matrices of $\bm{A}_i$ are assumed to be constant and are denoted by the $N \times N$ matrix $\bar{A}$ and the $N^2 \times N^2$ matrix $C_A$, respectively,
\begin{align}
\label{eqn:randcombinemean}
\bar{A} & \defeq \E\,\bm{A}_i = \begin{bmatrix}
\bar{a}_{11} & \bar{a}_{12} & \cdots & \bar{a}_{1N} \\
\bar{a}_{21} & \bar{a}_{22} & \cdots & \bar{a}_{2N} \\
\vdots & \vdots & \ddots & \vdots \\
\bar{a}_{N1} & \bar{a}_{N2} & \cdots & \bar{a}_{NN} \\
\end{bmatrix}  \\
\label{eqn:randcombinemeanelement}
\bar{a}_{\ell k} & \defeq \E\,\bm{a}_{\ell k}(i) \\
\label{eqn:randcombinecov}
C_A & \defeq \E [(\bm{A}_i - \bar{A}) \kron (\bm{A}_i - \bar{A})] = \begin{bmatrix}
C_{a,11} & C_{a,12} & \cdots & C_{a,1N} \\
C_{a,21} & C_{a,22} & \cdots & C_{a,2N} \\
\vdots & \vdots & \ddots & \vdots \\
C_{a,N1} & C_{a,N2} & \cdots & C_{a,NN} \\
\end{bmatrix}  \\
\label{eqn:randcombinecovblock}
C_{a,\ell k} & \defeq \E[(\bm{a}_{\ell k}(i) - \bar{a}_{\ell k})(\bm{A}_i - \bar{A})] = \begin{bmatrix}
c_{a,\ell k, 11} & c_{a,\ell k, 12} & \cdots & c_{a,\ell k, 1N} \\
c_{a,\ell k, 21} & c_{a,\ell k, 22} & \cdots & c_{a,\ell k, 2N} \\
\vdots & \vdots & \ddots & \vdots \\
c_{a,\ell k, N1} & c_{a,\ell k, N2} & \cdots & c_{a,\ell k, NN} \\
\end{bmatrix} 
\end{align}
where $\bar{a}_{\ell k}$ denotes the $(\ell,k)$-th element of $\bar{A}$, $C_{a,\ell k}$ denotes the $(\ell,k)$-th block with size $N\times N$ of $C_A$, and $c_{a,\ell k,nm}$ denotes the $(n,m)$-th element of $C_{a,\ell k}$. The scalar $c_{a,\ell k,nm}$ represents the covariance between the combination coefficients $\bm{a}_{\ell k}(i)$ and $\bm{a}_{nm}(i)$:
\be
\label{eqn:randcombinecoventry}
c_{a,\ell k,nm} \defeq \E [(\bm{a}_{\ell k}(i) - \bar{a}_{\ell k})(\bm{a}_{nm}(i) - \bar{a}_{nm})]
\ee
When $\ell = n$ and $k = m$, the scalar $c_{a,\ell k,\ell k}$ becomes the variance of $\bm{a}_{\ell k}(i)$.

\item The random matrices $\bm{M}_i$ and $\bm{A}_i$ are mutually-independent and are independent of any other random variable. 

\item We refer to the topology that corresponds to the average combination matrix $\bar{A}$ as the \emph{mean} graph, which is fixed over time. For each agent $k$, the neighborhood defined by the mean graph is denoted by $\Ncal_k$. The mean combination coefficients $\bar{a}_{\ell k} > 0$ satisfy the following constraints (compare with \eqref{eqn:topologyconstraints} and \eqref{eqn:randomtopologyconstraints}):
\be
\sum_{\ell \in \Ncal_k} \bar{a}_{\ell k} = 1, \;\;\mbox{and}\;\;
\begin{cases}
\bar{a}_{\ell k} > 0, & \mbox{if} \; \ell\in \Ncal_k \\
\bar{a}_{\ell k} = 0, & \mbox{otherwise} \\
\end{cases}
\ee
\end{enumerate}
The asynchronous network model described above is general enough to cover many situations of practical interest --- note that the model does not impose any specific probabilistic distribution on the step-sizes, network topologies, or combination coefficients. The upper bounds $\{ \mu_k \}$ are arbitrary and are independent of the constant step-size parameters used in synchronous diffusion networks \eqref{eqn:atcsync1}--\eqref{eqn:atcsync2}. For example, we can choose the sample space of each step-size ${\bm{\mu}}_k(i)$ to be the binary choice $\{0, \mu\}$ to model a random ``on-off'' behavior at each agent $k$ for the purpose of saving power, waiting for data, or even due to random agent failures. Similarly, we can choose the sample space of each combination coefficient ${\bm{a}}_{\ell\,k}(i)$, $\ell \in \Nknk$, to be $\{0, a_{\ell k}\}$ to model a random ``on-off'' status for the link from agent $\ell$ to agent $k$ at time $i$ for the purpose of either saving communication cost or due to random link failures. If links are randomly chosen by agents such that at every time $i$ there is only one other neighboring agent being communicated with, then we effectively mimic the random gossip strategies \cite{Boyd06TIT, Aysal09Allerton, Aysal09TSP, Jakovetic11TSP, Kar11TSP}. Note that the convex constraint \eqref{eqn:randomtopologyconstraints} can be satisfied by adjusting the value of $\bm{a}_{kk}(i)$ according to the realizations of $\{\bm{a}_{\ell k}(i); \ell \in \bm{\Ncal}_{k,i} \backslash \{k\}\}$. If the underlying topology is changing over time and the combination weights are also selected in a random manner, then we obtain the probabilistic diffusion strategy studied in \cite{Lopes08ICASSP, Takahashi10ICASSP} or the random link or topology model studied in \cite{Kar08TSP, Kar09TSP, Kar10TSP}. Since the parameter matrices $\bm{M}_i$ and $\bm{A}_i$ are assumed to be independent of each other and of any other random variable, the statistical dependency among the random variables $\{\bm{w}_i, \bm{\psi}_i, \bm{A}_i, \bm{M}_i\}$ is illustrated in Fig.~\ref{fig:Markov_asyn}. The filtration $\F_{i-1}$ now also includes information about $\bm{A}_{i-1}$ and $\bm{M}_{i-1}$ to represent \emph{all} information available up to iteration $i-1$.

\subsection{Properties of the Asynchronous Model}
\begin{figure}
\includegraphics[scale=0.5]{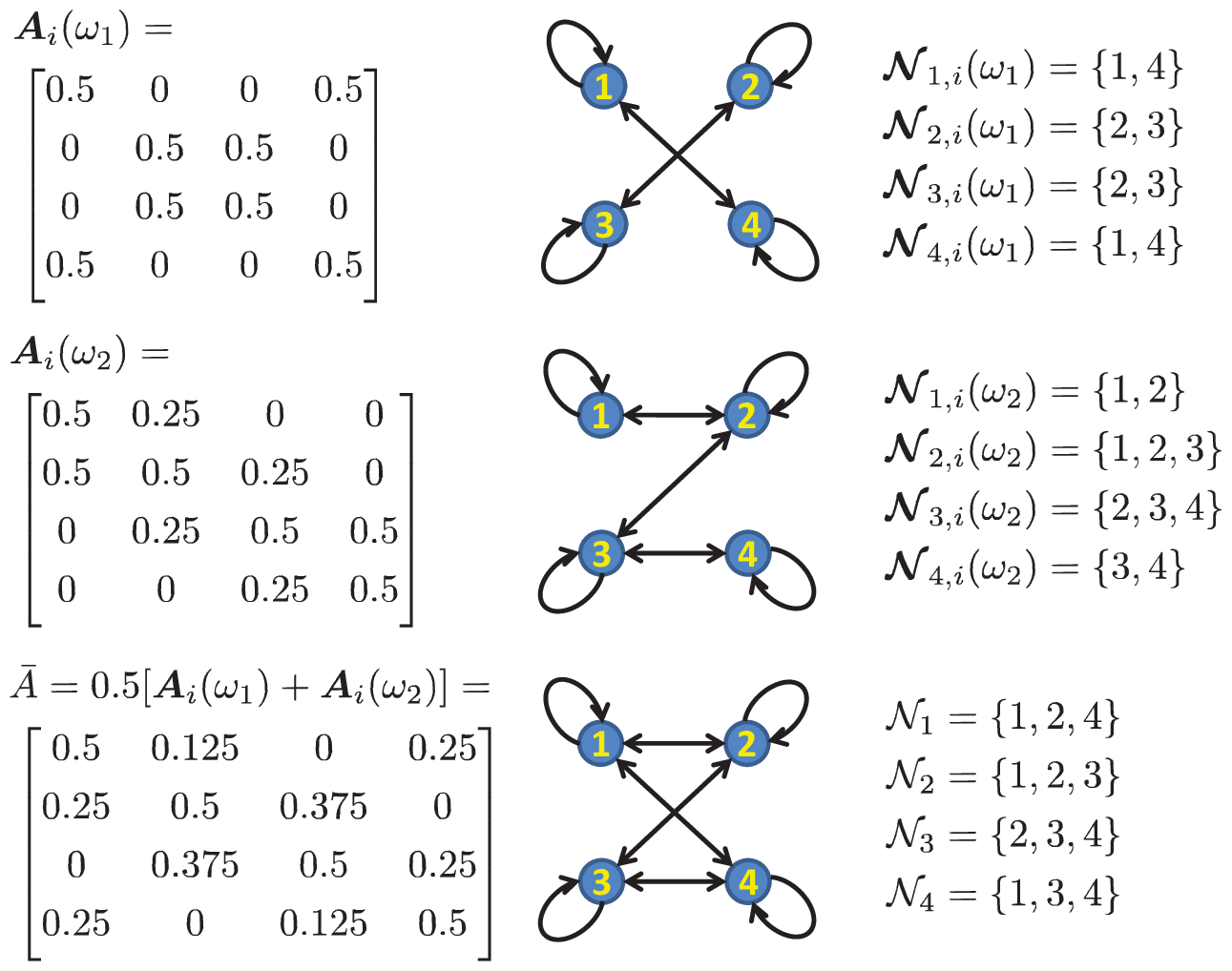}
\centering
\caption{The first two rows show two equally probable realizations with the respective neighborhoods. The last row shows the resulting mean graph.}
\label{fig:randomtopology}
\vspace{-1\baselineskip}
\end{figure}

The randomness of the combination coefficient matrix $\bm{A}_i$ arises from three sources. The first source is the randomness in the topology. The random topology is used to model the rich dynamics of evolving adaptive networks. The second source arises once a certain topology is realized, where the links among agents are further allowed to drop randomly. This phenomenon may be caused by either random interference or fading that blocks the communication links, or by neighbor selection policies used to save resources such as energy and bandwidth. The third sources relates to the agents which are allowed to assign random combination coefficients to their links, as long as the constraint \eqref{eqn:randomtopologyconstraints} is satisfied. An example of a random network with two equally probable realizations and its mean graph is shown in Fig.~\ref{fig:randomtopology}. The letter $\omega$ is used to index the sample space of the random matrix $\bm{A}_i$. A useful result relating the random neighborhoods $\{ \bm{\Ncal}_{k,i} \}$ from \eqref{eqn:atcasync2} to the neighborhoods $\{ \Ncal_k \}$ from the mean network model is given in the following statement.

\begin{lemma}[Neighborhoods]
\label{lemma:neighborhoods}
The neighborhood $\Ncal_k$ defined by the mean graph of the asynchronous network model is equal to the \emph{union} of all possible realizations for the random neighborhood $\bm{\Ncal}_{k,i}$ in \eqref{eqn:atcasync2}, i.e.,
\be
\Ncal_k = \bigcup_{\omega\in\Omega} \bm{\Ncal}_{k,i}(\omega)
\ee
for any $k$, where $\Omega$ denotes the sample space of $\bm{\Ncal}_{k,i}$.
\end{lemma}
\begin{IEEEproof}
We first establish $\bigcup_{\omega\in\Omega} \bm{\Ncal}_{k,i}(\omega) \subseteq \Ncal_k$. By \eqref{eqn:randomtopologyconstraints}, we have $\bm{a}_{\ell k}(i) > 0$ for any $\ell\in\bm{\Ncal}_{k,i}$. Since $\bm{a}_{\ell k}(i)$ is a nonnegative random variable, if the event $\bm{a}_{\ell k}(i) > 0$ occurs, then $\bar{a}_{\ell k} > 0$ by \eqref{eqn:randcombinemeanelement}, which implies $\ell \in {\Ncal}_k$. Thus, we get $\bm{\Ncal}_{k,i} \subseteq {\Ncal}_k$. This relation holds for any realization of $\bm{\Ncal}_{k,i}$, so we have $\bigcup_{\omega\in\Omega} \bm{\Ncal}_{k,i}(\omega) \subseteq \Ncal_k$.

\begin{figure}
\includegraphics[scale=0.3]{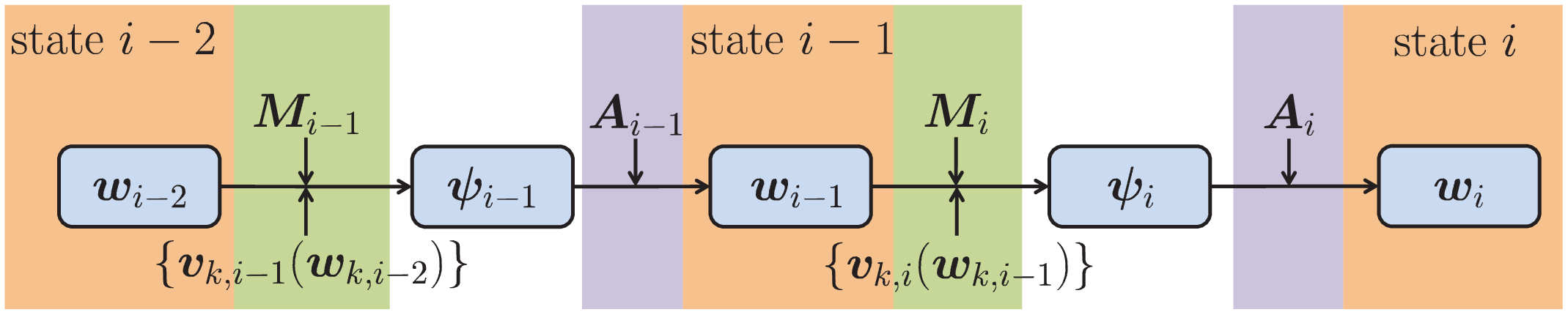}
\centering
\caption{An illustration of the statistical dependency for the asynchronous diffusion strategy \eqref{eqn:atcasync1}--\eqref{eqn:atcasync2}}
\label{fig:Markov_asyn}
\vspace{-1\baselineskip}
\end{figure}

Now we establish $\Ncal_k \subseteq \bigcup_{\omega\in\Omega} \bm{\Ncal}_{k,i}(\omega)$. For any $\ell \in {\Ncal}_k$, we have $\bar{a}_{\ell k} > 0$ by definition. This is only possible if there exists at least one realization of $\bm{a}_{\ell k}(i)$ assuming a positive value, which means that $\ell \in \bm{\Ncal}_{k,i}(\omega)$ for a certain $\omega$. Therefore, $\Ncal_k \subseteq \bigcup_{\omega\in\Omega} \bm{\Ncal}_{k,i}(\omega)$ holds as expected.
\end{IEEEproof}

Another useful property for the asynchronous model relates to the  combination coefficient matrices $\{\bar{A}, \bar{A} \kron \bar{A} + C_A\}$.

\begin{lemma}[Left-stochastic matrices]
\label{lemma:leftstochastic}
The $N \times N$ matrix $\bar{A}$ and the $N^2\times N^2$ matrix $\bar{A} \kron \bar{A} + C_A$ are left-stochastic matrices, meaning that every element of $\bar{A}$ or $\bar{A} \kron \bar{A} + C_A$ is nonnegative and 
\be
\label{eqn:AACAone}
\bar{A}^\T \one_N = \one_N, \qquad (\bar{A} \kron \bar{A} + C_A)^\T \one_{N^2} = \one_{N^2}
\ee
\end{lemma}
\begin{IEEEproof}
Since $\bm{A}_i$ has nonnegative entries by the asynchronous network model, it is easy to verify that $\bar{A}$ and $\bm{A}_i \kron \bm{A}_i$ also have nonnegative entries. Moreover, noting that
\begin{align}
\label{eqn:AkApCA}
\E(\bm{A}_i \kron \bm{A}_i) & = \bar{A} \kron \bar{A} + \E[(\bm{A}_i - \bar{A}) \kron (\bm{A}_i - \bar{A})] \nn \\
& = \bar{A} \kron \bar{A} + C_A
\end{align}
it follows that $\bar{A} \kron \bar{A} + C_A$ has nonnegative entries as well. Furthermore, observe that
\be
\bar{A}^\T \one_N = \E(\bm{A}_i^\T) \one_N = \E(\bm{A}_i^\T \one_N) = \one_N
\ee
and
\begin{align}
(\bar{A} \kron \bar{A} + C_A)^\T \one_{N^2} & = \E(\bm{A}_i^\T \kron \bm{A}_i^\T ) (\one_N \kron \one_N) \nn \\
& = \E[(\bm{A}_i^\T \one_N) \kron (\bm{A}_i^\T \one_N )] \nn \\
& = \one_{N^2}
\end{align}
as desired.
\end{IEEEproof}

A useful special case of the asynchronous network model is the spatially-uncorrelated model, where the random step-sizes at the agents are uncorrelated with each other across the network, and the random combination coefficients assigned by each agent to its local neighbors \emph{excluding} itself are also uncorrelated with each other and with all other combinations weights assigned by other agents across the network. In the next subsection we provide two concrete examples for this model.

\begin{lemma}[The spatially-uncorrelated model]
\label{lemma:spatial}
Under the asynchronous network model, if at each iteration $i$, the random step-sizes $\{\bm{\mu}_k(i); k = 1,2,\dots, N\}$ are uncorrelated with each other across the network, and if the random combination coefficients $\{\bm{a}_{\ell k}(i); \ell  \ne  k, k=1,2,\dots,N\}$ are also assumed to be uncorrelated with each other across the network, then the covariances $\{c_{\mu,k,\ell}\}$ in \eqref{eqn:randstepsizecoventry} and $\{c_{a,\ell k, nm}\}$ in \eqref{eqn:randcombinecoventry} are now given by
\begin{align}
\label{eqn:cov_mu_kl}
c_{\mu,k,\ell} & = \begin{cases}
c_{\mu,k,k}, & \mbox{if} \; \ell = k \\
0, & \mbox{otherwise} \\
\end{cases} \\
\label{eqn:cov_a_lk_nm}
c_{a,\ell k, nm} & = \begin{cases}
c_{a,\ell k, \ell k}, & \mbox{if} \; k  =  m, \ell  =  n, \ell \in \Nknk \\
-c_{a,\ell k, \ell k}, & \mbox{if} \; k  =  m  =  n, \ell \in \Nknk \\
-c_{a,nk, nk}, & \mbox{if} \; k  =  m  =  \ell, n \in \Nknk \\
\displaystyle \sum_{j \in\Nknk} c_{a,j k, jk}, & \mbox{if} \; k  =  m  =  \ell  =  n \\
0, & \mbox{otherwise} \\
\end{cases}
\end{align}
Correspondingly, the block matrices $\{C_{\mu,k}, C_{a,\ell k}\}$ in \eqref{eqn:randstepsizecovblock} and \eqref{eqn:randcombinecovblock} are given by the following compact expressions:
\begin{align}
\label{eqn:cov_mat_mu_kl}
C_{\mu,k} & = c_{\mu,k,k} \cdot E_{kk} \\
\label{eqn:cov_mat_a_lk_nm}
C_{a,\ell k} & = c_{a,\ell k, \ell k} \cdot (E_{\ell k} - E_{kk}), \;\; \ell \in \Nknk \\
\label{eqn:cov_mat_a_kk_nm}
C_{a,kk} & = \sum_{\ell \in\Nknk} c_{a,\ell k, \ell k} \cdot (E_{kk} - E_{\ell k})
\end{align}
where $E_{\ell k}$ denotes the $N\times N$ matrix whose entries are all zero except for the $(\ell,k)$-th entry, which is equal to one.
\end{lemma}
\begin{IEEEproof}
See Appendix \ref{app:spatial}.
\end{IEEEproof}

\noindent We remark that the matrices $\{C_M,C_A\}$ are Kronecker-covariance matrices defined by \eqref{eqn:randstepsizecov} and \eqref{eqn:randcombinecov}; they are \emph{not} conventional covariance matrices and, therefore, are \emph{not} necessarily Hermitian matrices.

\subsection{Two Useful Network Models}
\label{subsec:twoexample_model}
In this subsection we describe two scenarios where the asynchronous behavior arises naturally. The first model below is referred to as the Bernoulli model, a special case of which was used before to model random link failures over consensus networks \cite{Kar09TSP, Kar10TSP}. The Bernoulli model given here is more general in that it also allows us to consider simultaneously on-off strategies for adaptation through equation \eqref{eqn:binary_randomstepsize}.

\subsubsection{The Bernoulli Model}
\label{subsubsec:bernoulli}
We assume that at every time $i$, each agent $k$ adopts a random ``on-off'' policy to reduce energy consumption. Specifically, agent $k$ enters an active mode with probability $0 < q_k < 1$ and performs the self-learning step \eqref{eqn:atcasync1}, and it enters a sleep mode with probability $1-q_k$ to save energy. This behavior can also be interpreted as the result of random data arrivals: at every time $i$, a new data becomes available to agent $k$ with probability $q_k$. This situation can be modeled by the following Bernoulli random step-size model:
\be
\label{eqn:binary_randomstepsize}
\bm{\mu}_k(i) = \begin{cases}
\mu_{k}, & \mbox{with probability} \;\; q_k \\
0 , & \mbox{with probability} \;\; 1-q_k \\
\end{cases}
\ee
where $\mu_{k}$ is a fixed step-size. We further assume that the underlying topology is fixed. However, each agent $k$ is allowed to randomly choose a \emph{subset} of its neighbors to perform the social-learning step \eqref{eqn:atcasync2}. Specifically, agent $k$ chooses neighbor $\ell$ with probability $0 < \eta_{\ell k} < 1$ to save on communication costs. This behavior can also be interpreted as resulting from random link failures: at every time $i$, the communication link from agent $\ell$ to agent $k$ drops with probability $\eta_{\ell k}$. This situation can be modeled by the following Bernoulli random combination coefficients model:
\be
\label{eqn:binary_randomcombine}
\bm{a}_{\ell k}(i) = \begin{cases}
a_{\ell k}, & \mbox{with probability} \;\; \eta_{\ell k} \\
0, & \mbox{with probability} \;\; 1-\eta_{\ell k} \\
\end{cases}
\ee
for any $\ell  \in  \bm{\Ncal}_{k,i}\backslash\{k\}$, where $0  <  a_{\ell k}  <  1$ is a fixed combination coefficient. Based on Lemma \ref{lemma:neighborhoods}, we require the values of $\bm{a}_{\ell k}(i)$ in \eqref{eqn:binary_randomcombine} to ensure that $0 \le \bm{a}_{kk}(i) \le 1$ where
\be
\label{eqn:akkibounded}
\bm{a}_{kk}(i) = 1 - \sum_{\ell \in \bm{\Ncal}_{k,i}\backslash\{k\}} \bm{a}_{\ell k}(i)
\ee
Using Lemma \ref{lemma:spatial}, the relevant quantities introduced in the asynchronous network model are given by
\begin{align}
\label{eqn:bernoulli_M_mean}
\bar{\mu}_k & = q_k\mu_k \\
\label{eqn:bernoulli_M_cov}
c_{\mu,k,k} & = q_k(1-q_k)\mu_k^2 \\
\label{eqn:bernoulli_A_mean}
\bar{a}_{\ell k} & = \eta_{\ell k} a_{\ell k} \\
\label{eqn:bernoulli_A_mean_diag}
\bar{a}_{kk} & \defeq 1 - \sum_{\ell \in \Nknk} \eta_{\ell k}a_{\ell k} \\
\label{eqn:bernoulli_A_cov_lk}
c_{a,\ell k, \ell k} & = \eta_{\ell k}(1-\eta_{\ell k})a_{\ell k}^2, \quad  \ell \in \Nknk
\end{align}

\subsubsection{The Beta Model}
\begin{figure}
\includegraphics[scale=0.45]{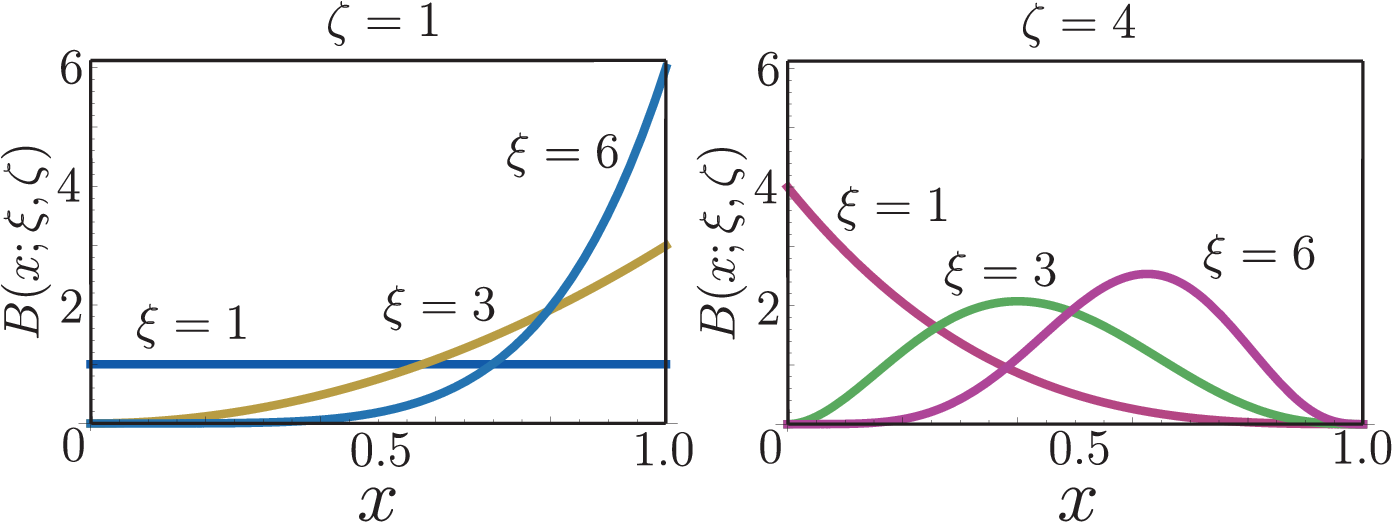}
\centering
\caption{The PDFs of the Beta distribution $B(x; \xi, \zeta)$ for different values of $\xi$ and $\zeta$.}
\label{fig:beta}
\vspace{-1\baselineskip}
\end{figure}

The other example involves continuous random variables modeled by Beta distributions, which can be viewed as extensions of binary Bernoulli distributions to the continuous domain when the probability mass is distributed over a bounded region. The family of Beta distributions takes values in the interval $[0,1]$ and includes the uniform distribution over $[0,1]$ as a special case \cite{DeGroot11}. The probability density function (PDF) of a Beta distribution is given by
\be
\label{eqn:standard_beta}
B(x; \xi, \zeta) = \begin{cases}
\displaystyle \frac{\Gamma(\xi + \zeta)}{\Gamma(\xi)\Gamma(\zeta)}x^{\xi - 1}(1  -  x)^{\zeta - 1},   &  0  \le  x  \le  1 \\
0,  &  \mbox{otherwise} \\
\end{cases}
\ee
where $\xi, \zeta > 0$ are the shape parameters and $\Gamma(\cdot)$ denotes the Gamma function. Figure \ref{fig:beta} plots $B(x; \xi, \zeta)$ for two values of $\zeta$. The mean and variance of the Beta distribution \eqref{eqn:standard_beta} are given by
\be
\label{eqn:beta_meanandvariance}
\bar{x} = \frac{\xi}{\xi + \zeta}, \quad \sigma_x^2 = \frac{\xi \zeta}{(\xi + \zeta)^2(\xi + \zeta + 1)}
\ee

For the asynchronous network model, we assume that the step-size $\bm{\mu}_k(i)$ takes random values in the range $[0, \mu_k]$, where $\mu_k$ denotes the largest possible value for the $k$-th step-size. We further assume that the scaled parameter 
$\bm{\mu}_k(i) / \mu_k$ is governed by a Beta distribution:
\be
\label{eqn:beta_randomstepsize}
\bm{x}_k(i) = \frac{\bm{\mu}_k(i)}{\mu_k} \sim B(x_k; \xi_k, \zeta_k)
\ee
where $\{\xi_k, \zeta_k >0\}$ are the corresponding shape parameters. Likewise, we assume that the combination coefficient $\bm{a}_{\ell k}(i)$ for $\ell \in \bm{\Ncal}_{k,i}\backslash\{k\}$ takes random values in the range $[0, a_{\ell k}]$ with $0  <  a_{\ell k}  <  1$. The scaled parameter $\bm{a}_{\ell k}(i) / a_{\ell k}$ is assumed to be governed by a Beta distribution:
\be
\label{eqn:beta_randomcombine}
\bm{y}_{\ell k}(i) = \frac{\bm{a}_{\ell k}(i)}{a_{\ell k}} \sim B(y_{\ell k}; \xi_{\ell k}, \zeta_{\ell k})
\ee
where $\{\xi_{\ell k}, \zeta_{\ell k} > 0\}$ are the shape parameters. We adopt the spatially uncorrelated model from Lemma \ref{lemma:spatial}. In order to guarantee that $\bm{a}_{k k}(i)$ always assumes values within the range $[0,1]$, we again require condition \eqref{eqn:akkibounded}. Then, we can use \eqref{eqn:beta_meanandvariance} to calculate the relevant quantities introduced in the asynchronous network model:
\begin{align}
\label{eqn:beta_M_mean}
\bar{\mu}_k & = \frac{\xi_k}{\xi_k + \zeta_k} \mu_k \\
\label{eqn:beta_M_cov}
c_{\mu,k,k} & = \frac{\xi_k \zeta_k}{(\xi_k + \zeta_k)^2(\xi_k + \zeta_k + 1)} \mu_k^2 \\
\label{eqn:beta_A_mean}
\bar{a}_{\ell k} & = \frac{\xi_{\ell k}}{\xi_{\ell k} + \zeta_{\ell k}} a_{\ell k} \\
\label{eqn:beta_A_mean_diag}
\bar{a}_{kk} & \defeq 1 - \sum_{\ell \in \Ncal\backslash\{k\}} \frac{\xi_{\ell k}}{\xi_{\ell k} + \zeta_{\ell k}} a_{\ell k} \\
\label{eqn:beta_A_cov_lk}
c_{a,\ell k, \ell k} & = \frac{\xi_{\ell k} \zeta_{\ell k}}{(\xi_{\ell k}  +  \zeta_{\ell k})^2(\xi_{\ell k}  +  \zeta_{\ell k}  +  1)} a_{\ell k}^2, \;  \ell  \in  \Nknk
\end{align}

\section{Main Results}
We presented a fairly general model for asynchronous events to cover many useful scenarios. Due to the random evolution of the step-sizes, topology, neighborhoods, and combination coefficients, the error dynamics of the network is influenced by more factors than traditional synchronous networks. Hence, before delving into the technical details, we summarize in this section some of the main conclusions that will follow from the analysis for the benefit of the reader.

The first important result (in Part I) is Theorem \ref{theorem:meansquarestability}, which presents in \eqref{eqn:meansquarestabilitycond3} a condition on the first and second-order moments of the random step-sizes to ensure the mean-square stability of the asynchronous network. It is noteworthy and also remarkable that the random distribution of the combination coefficients (and, hence, the randomness in the topology) does not affect the stability condition. This is another manifestation of the property stated earlier in the introduction, and which has been proven before only for synchronous networks in \cite{Chen11TSP}, that the mean-square stability of diffusion strategies is insensitive to the network topology even in the asynchronous case. Theorem \ref{theorem:meansquarestability} will show that studying the mean-square stability of the asynchronous network can be achieved by investigating the stability of the following recursive inequality:
\be
\label{eqn:worstMSDrecursion}
\max_k \E\| \wt{\bm{w}}_{k,i}\|^2 \le \beta \cdot \max_k \E\| \wt{\bm{w}}_{k,i-1}\|^2 + \theta \sigma_v^2
\ee
where $\wt{\bm{w}}_{k,i} = w^o - \bm{w}_{k,i}$ denotes the error vector at agent $k$ at time $i$, and $\{\beta, \theta, \sigma_v^2 \}$ are certain parameters defined by \eqref{eqn:betadef}, \eqref{eqn:thetadef}, and \eqref{eqn:vkicond_var}, respectively. The recursive inequality \eqref{eqn:worstMSDrecursion} provides an upper bound for the \emph{worst} individual MSD in the network at every time $i$. Thus, as long as the factor $\beta$ is inside the unit circle, i.e., $|\beta| < 1$, then the sequences $\{ \E\| \wt{\bm{w}}_{k,i}\|^2; i\ge0 \}$ for all $k$ will remain bounded and the network will be mean-square stable. Theorem \ref{theorem:meansquarestability} provides a sufficient condition to guarantee $|\beta| < 1$. The condition involves the first and second-order moments of the step-size distribution as follows:
\be
\label{eqn:conditionmeansquarestable}
\frac{\bar{\mu}_k^2 + c_{\mu,k,k}}{\bar{\mu}_k} < \frac{\lambda_{k,\min}}{\alpha + \lambda_{k,\max}^2}
\ee
for all $k$, where $\{\lambda_{k,\min}, \lambda_{k,\max}\}$ and $\alpha$ are from Assumptions \ref{asm:boundedHessian} and \ref{asm:gradientnoise}, respectively. As long as condition \eqref{eqn:conditionmeansquarestable} holds, the random step-sizes will lead to stable networks \emph{regardless} of their PDFs. Furthermore, when condition \eqref{eqn:conditionmeansquarestable} holds, the same theorem establishes that the MSD of the individual agents will eventually be bounded by
\be
\label{eqn:worseMSDbound}
\limsup_{i\rightarrow\infty}\E\|\wt{\bm{w}}_{k,i}\|^2 \le  b \cdot \nu_o
\ee
where $b$ and $\nu_o$ are given by
\eqref{eqn:mumaxdef} and $\nu_o$ is reproduced here:
\be
\nu_o = \max_{k} \frac{\bar{\mu}_k^2 + c_{\mu,k,k}}{\bar{\mu}_k} = \frac{ \bar{\mu}_k^{(2)} }{\bar{\mu}_k^{(1)}}
\ee
where $\bar{\mu}_k^{(m)} \defeq \E [\bm{\mu}_k(i)]^m $ denotes the $m$-th moment of $\bm{\mu}_k(i)$ and $\bar{\mu}_k \equiv \bar{\mu}_k^{(1)}$. Thus, the individual MSD values will decrease with $\nu_o$ and can become arbitrarily small in steady-state. 

\begin{figure}
\includegraphics[scale=1]{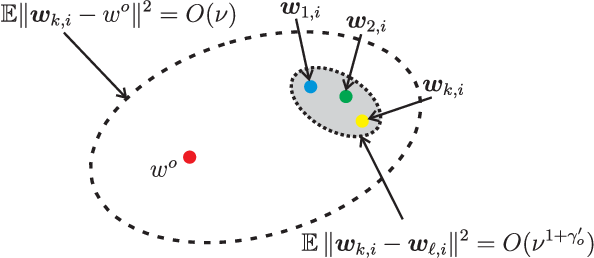}
\centering
\caption{Illustration for results \eqref{eqn:individualMSDapproxOnu} and \eqref{eqn:crossMSDapproxOnu2}: the solutions by the agents do not only get $O(\nu)$ close to the target $w^o$ but they also cluster next to each other within $O(\nu^{1 + \gamma_o'})$ for some $\gamma_o' > 0$.}
\label{fig:normball}
\vspace{-1\baselineskip}
\end{figure}

With some strengthened assumptions on the gradient noise and random step-sizes, and using Theorem \ref{theorem:meansquarestability}, we then establish in Part II \cite{Zhao13TSPasync2} two important conclusions:
\begin{align}
\label{eqn:individualMSDapproxOnu}
\lim_{i\rightarrow \infty} \E\,\|w^o - \bm{w}_{k,i}\|^2 & = O(\nu) \\
\label{eqn:crossMSDapproxOnu2}
\lim_{i\rightarrow \infty} \E\,\|\bm{w}_{k,i} - \bm{w}_{\ell,i}\|^2 & = O(\nu^{1 + \gamma_o'})
\end{align}
for some $\gamma_o' > 0$ and sufficiently small $\nu$. The parameter $\nu$ is defined by
\be
\nu \defeq \max_{k} \frac{ \sqrt{ \bar{\mu}_k^{(4)} }}{\bar{\mu}_k^{(1)}}
\ee
These results imply that all agents reach a level of $O(\nu^{1 + \gamma_o'})$ agreement with each other and get $O(\nu)$ close to the desired solution $w^o$ in steady-state. This interesting behavior is illustrated in Fig.~\ref{fig:normball}, where it is shown that, despite being subjected to different sources of randomness and failures, the agents in an asynchronous network are still able to approach the desired solution and they are also able to coalesce close to each other while seeking the desired solution. Actually, in Part II \cite{Zhao13TSPasync2}, we shall derive explicit closed-form expressions for the size of the error variances in \eqref{eqn:individualMSDapproxOnu} and \eqref{eqn:crossMSDapproxOnu2}.

\section{Mean-Square Stability}
\label{sec:meansquarestability}
For each agent $k$, we introduce the error vectors:
\be
\label{eqn:werrordef}
\wt{\bm\psi}_{k,i} \defeq w^o - \bm\psi_{k,i}, \qquad \wt{\bm{w}}_{k,i} \defeq w^o - \bm{w}_{k,i}
\ee
where $w^o$ is the desired optimal solution. Subtracting $w^o$ from both sides of \eqref{eqn:atcsync1}--\eqref{eqn:atcsync2} and using \eqref{eqn:linearperturbationmodel} gives
\begin{subequations}
\begin{align}
\label{eqn:atcasync1error}
\wt{\bm{\psi}}_{k,i} & = \wt{\bm{w}}_{k,i - 1}  +  \bm{\mu}_k(i)[\nabla_{w^*}J_k(\bm{w}_{k,i - 1})  +  \bm{v}_{k,i}(\bm{w}_{k,i - 1})] \\
\label{eqn:atcasync2error}
\wt{\bm{w}}_{k,i} & = \sum_{\ell\in\bm{\Ncal}_{k,i}} \bm{a}_{\ell k}(i)\,\wt{\bm{\psi}}_{\ell,i}
\end{align}
\end{subequations}
Applying the transformation $\ubar\mbbT(\cdot)$ from \eqref{eqn:ubardef} to both sides of these equations, we show in Appendix \ref{app:errorrecursion} that the error recursion \eqref{eqn:atcasync1error}--\eqref{eqn:atcasync2error} becomes
\begin{subequations}
\begin{align}
\label{eqn:atcasync1errorubar1}
\ubar{\wt{\bm{\psi}}}_{k,i} &  =  [I_{2M} - \bm{\mu}_k(i)\bm{H}_{k,i - 1}]\ubar{\wt{\bm{w}}}_{k,i - 1}  +  \bm{\mu}_k(i)\ubar{\bm{v}}_{k,i}(\bm{w}_{k,i - 1}) \\
\label{eqn:atcasync2errorubar1}
\ubar{\wt{\bm{w}}}_{k,i} &  =  \sum_{\ell\in\bm{\Ncal}_{k,i}} \bm{a}_{\ell k}(i)\,\ubar{\wt{\bm{\psi}}}_{\ell,i}
\end{align}
\end{subequations}
where we introduced the $2M\times2M$ matrix:
\be
\label{eqn:Hkidef}
\bm{H}_{k,i-1} \defeq \int_{0}^{1}\nabla_{\ubar{w}\ubar{w}^*}^2 J_k(\ubar{w}^o-t\ubar{\wt{\bm{w}}}_{k,i-1})\,dt
\ee
To proceed, we introduce the following network variables:
\begin{align}
\label{eqn:networkerrorvectordef}
\ubar{\wt{\bm{w}}}_i & \defeq \col\{\ubar{\wt{\bm{w}}}_{1,i},\ubar{\wt{\bm{w}}}_{2,i},\dots,\ubar{\wt{\bm{w}}}_{N,i}\} \\
\label{eqn:bigMidef}
\bm{\Mcal}_i & \defeq \bm{M}_i \kron I_{2M} \\
\label{eqn:bigAidef}
\bm{\Acal}_i & \defeq \bm{A}_i \kron I_{2M} \\
\bm{\Hcal}_i & \defeq \diag\{\bm{H}_{1,i},\bm{H}_{2,i},\dots,\bm{H}_{N,i}\}
\end{align}
Using \eqref{eqn:atcasync1errorubar1}--\eqref{eqn:atcasync2errorubar1} we conclude that the network error vector \eqref{eqn:networkerrorvectordef} evolves according to the following dynamics:
\be
\label{eqn:errorrecursiondef}
\boxed{
\ubar{\wt{\bm{w}}}_i = \bm{\Acal}_i^\T (I_{2MN}  -  \bm{\Mcal}_i \bm{\Hcal}_{i - 1})\ubar{\wt{\bm{w}}}_{i - 1}  +  \bm{\Acal}_i^\T\bm{\Mcal}_i\ubar{\bm{v}}_i(\bm{w}_{i - 1})
}
\ee

\subsection{Condition for Mean-Square Stability}

The main result in this Part I is to establish the mean-square stability of this recursion in Theorem \ref{theorem:meansquarestability} below. The difficulty lies in the fact that the error dynamics \eqref{eqn:errorrecursiondef} is a time-variant stochastic recursion that also depends nonlinearly on the data. The parameters $\{\bm{\Acal}_i, \bm{\Mcal}_i, \bm{\Hcal}_{i-1}\}$ are random, time-varying, and multiplied together and by the error vector and noise variables. The statement and proof of Theorem \ref{theorem:meansquarestability} rely on the following quantities:
\begin{align}
\label{eqn:epsilondef}
\epsilon^2(i) & \defeq \max_k \E\| \wt{\bm{w}}_{k,i}\|^2 = \frac{1}{2} \cdot \max_k \E\|\ubar{\wt{\bm{w}}}_{k,i}\|^2 \\
\label{eqn:gammak2def}
\gamma_k^2 & \defeq 1 - 2\bar{\mu}_k\lambda_{k,\min} + (\bar{\mu}_k^2 + c_{\mu,k,k}) \lambda_{k,\max}^2 \\
\label{eqn:betadef}
\beta & \defeq \max_k\{\gamma_k^2 + \alpha(\bar{\mu}_k^2 + c_{\mu,k,k} )\} \\
\label{eqn:thetadef}
\theta & \defeq \max_k\{\bar{\mu}_k^2 + c_{\mu,k,k} \}
\end{align}
where the $\{\lambda_{k,\min},\lambda_{k,\max}\}$ correspond to the lower and upper bounds on the Hessian matrices from \eqref{eqn:boundseigHessian}.

\begin{theorem}[Mean-square stability]
\label{theorem:meansquarestability}
The mean-square stability of the asynchronous diffusion strategy \eqref{eqn:atcasync1}--\eqref{eqn:atcasync2} reduces to studying the convergence of the recursive inequality:
\be
\label{eqn:meansquareerrorboundrecursion}
\epsilon^2(i) \le \beta \cdot \epsilon^2(i-1) + \theta\sigma_v^2
\ee
where $\sigma_v^2$ is from \eqref{eqn:vkicond_var}. The model \eqref{eqn:meansquareerrorboundrecursion} is stable if the mean $\{\bar{\mu}_k\}$ and the ratio $\{(\bar{\mu}_k^2 + c_{\mu,k,k})/\bar{\mu}_k\}$ satisfy the following relation:
\be
\label{eqn:meansquarestabilitycond3}
\boxed{
\frac{\bar{\mu}_k^2 + c_{\mu,k,k}}{\bar{\mu}_k} < \frac{\lambda_{k,\min}}{\alpha + \lambda_{k,\max}^2}
}
\ee
for $k=1,2,\dots, N$, where the parameter $\alpha$ is from \eqref{eqn:gradientnoise2norm}. When condition \eqref{eqn:meansquarestabilitycond3} holds, an upper bound on the individual steady-state mean-square-deviation (MSD) for each agent $k$ in the network is given by
\be
\label{eqn:meansquareerrorbound7}
\boxed{
\limsup_{i\rightarrow\infty} \E\|\wt{\bm{w}}_{k,i}\|^2 \le  b \cdot \nu_o
}
\ee
where
\be
\label{eqn:mumaxdef}
\nu_o \defeq \max_{k} \frac{\bar{\mu}_k^2 + c_{\mu,k,k}}{\bar{\mu}_k}, \quad
b \defeq \frac{\kappa\sigma_v^2}{\min_k\{\lambda_{k,\min}\}}
\ee
and the parameter $\kappa$ is from \eqref{eqn:kappadef}.
\end{theorem}
\begin{IEEEproof}
See Appendix \ref{app:proofofstablity}.
\end{IEEEproof}

From the asynchronous network model, we know that $\bm{\mu}_k(i) \in [0, \mu_k]$. It follows that
\be
\label{eqn:simplerbound}
\frac{\bar{\mu}_k^2 + c_{\mu,k,k}}{\bar{\mu}_k} = \frac{ \bar{\mu}_k^{(2)} }{\bar{\mu}_k^{(1)} } \le \frac{ \E\,[\bm{\mu}_k(i) \mu_k] }{\bar{\mu}_k} = \mu_k
\ee
From \eqref{eqn:simplerbound}, a sufficient condition for \eqref{eqn:meansquarestabilitycond3} to hold is given by
\be
\label{eqn:sufficientconditionstability}
\mu_k < \frac{\lambda_{k,\min}}{\alpha + \lambda_{k,\max}^2}
\ee

Condition \eqref{eqn:meansquarestabilitycond3} allows us to provide some insights about how the dispersion of $\bm{\mu}_k(i)$ affects mean-square stability. Note that condition \eqref{eqn:meansquarestabilitycond3} even allows the \emph{random} step-sizes to assume some abnormally large values at a relatively low probability. This ``hopping'' behavior (resulting from infrequent large step-sizes) would not destroy the mean-square stability of the network; this fact reveals another useful form of robustness.

Since the constant coefficient $b$ defined in \eqref{eqn:mumaxdef} is a fixed bound, Theorem \ref{theorem:meansquarestability}  implies that for sufficiently large $i$, the MSD of each individual agent's solution has a bounded value. The upper bound in \eqref{eqn:meansquareerrorbound7} is proportional to the parameter $\nu_o$ across the network. Using the useful conclusion of \eqref{eqn:meansquareerrorbound7}, we will be able to derive in the sequel a condition for fourth-order stability of the error recursion \eqref{eqn:errorrecursiondef}.

\subsection{Stability Conditions for Bernoulli and Beta Models}
We specialize condition \eqref{eqn:meansquareerrorbound7} for the asynchronous models described in Section \ref{subsec:twoexample_model}. 

\subsubsection{The Bernoulli Model}
Substituting \eqref{eqn:bernoulli_M_mean} and \eqref{eqn:bernoulli_M_cov} into \eqref{eqn:meansquarestabilitycond3} yields the condition
\be
\label{eqn:muk_bound_bernoulli}
\mu_k < \frac{\lambda_{k,\min}}{\alpha + \lambda_{k,\max}^2}
\ee
which is identical to condition \eqref{eqn:sufficientconditionstability} on the upper limit of the range of random step-sizes.

\subsubsection{The Beta Model}
\begin{figure}
\includegraphics[scale=0.7]{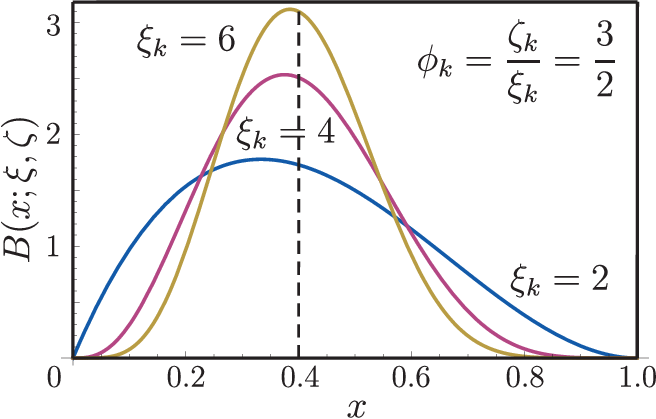}
\centering
\caption{The PDFs of the Beta distribution $B(x;\xi_k,\zeta_k)$ for $\zeta_k = 1.5 \xi_k$ and $\xi_k = 2, 4, 6$.}
\label{fig:beta_fix}
\vspace{-1\baselineskip}
\end{figure}

Without loss of generality, let $\zeta_k = \phi_k \cdot \xi_k$ with a constant factor $\phi_k > 0$. It follows from \eqref{eqn:beta_M_mean} that the mean value $\bar{\mu}_k$ can be expressed in terms of the factor $\phi_k$ and the upper limit $\mu_k$:
\be
\label{eqn:mean_beta_fix}
\bar{\mu}_k = \frac{\mu_k}{1 + \phi_k}
\ee
Likewise, from \eqref{eqn:beta_M_cov}, we have
\be
\label{eqn:cov_beta_fix}
c_{\mu,k,k} = \frac{\phi_k \mu_k^2}{(1 + \phi_k)^2(\xi_k + \xi_k \phi_k + 1)} 
\ee
which is a monotonically decreasing function of the shape parameter $\xi_k \ge 1$. As the value of $\xi_k$ becomes larger, the probability mass of $\bm{\mu}_k(i)$ will gradually concentrate around its mean \eqref{eqn:mean_beta_fix}, as shown in Fig.~\ref{fig:beta_fix}. Substituting \eqref{eqn:mean_beta_fix} and \eqref{eqn:cov_beta_fix} into \eqref{eqn:meansquarestabilitycond3} yields
\be
\label{eqn:muk_bound_beta}
\mu_k < \left( 1 + \frac{\phi_k \xi_k}{1 + \xi_k} \right)\frac{\lambda_{k,\min}}{\alpha + \lambda_{k,\max}^2}
\ee
where $\mu_k$ is the largest possible value for $\bm{\mu}_k(i)$ defined by \eqref{eqn:beta_randomstepsize}. In \eqref{eqn:muk_bound_beta}, the bound on $\mu_k$ is a monotonically increasing function of the shape parameter $\xi_k \ge 1$. As $\xi_k$ becomes larger, the bound in \eqref{eqn:muk_bound_beta} becomes larger. The net effect allows for a wider range for the realizations of the random step-sizes. Moreover, it is easy to verify that the upper bound in \eqref{eqn:muk_bound_beta} is larger than that in \eqref{eqn:muk_bound_bernoulli}.

\subsection{Condition for Fourth-Order Stability}
Result \eqref{eqn:meansquareerrorbound7} establishes that the network is mean-square stable under the assumption of bounded second-order moments for the gradient noise process as in \eqref{eqn:gradientnoise2norm}. If desired, under a similar condition on bounded fourth-order moments for the gradient noise, we can also establish by extending the arguments of Appendix \ref{app:proofofstablity} and \cite{Chen13TIT} that the error recursion \eqref{eqn:errorrecursiondef} is stable in the fourth-order sense.

\begin{theorem}[Stability of fourth-order error moments]
\label{theorem:4thmoments}
Assume the fourth-order moments of the gradient noise components are bounded by
\be
\label{eqn:gradientnoise4thmoment}
\E[ \|\bm{v}_{k,i}(\bm{w}_{k,i-1}) \|^4 | \F_{i-1} ] \le \alpha^2 \| w^o - \bm{w}_{k,i-1}\|^4 + \sigma_v^4
\ee
for some constants $\alpha \ge 0$ and $\sigma_v \ge 0$. If
\be
\label{eqn:boundstepsize4thorder}
\frac{\sqrt{ \bar{\mu}_k^{(4)} }}{ \bar{\mu}_k } < \frac{ \lambda_{k, \min} }{ 3\lambda_{k, \max}^2 + 4\alpha }
\ee
holds for all $k$, then the fourth-order moments of the individual errors are asymptotically bounded by
\be
\label{eqn:bounded4thorderErrors}
\boxed{
\limsup_{i\rightarrow\infty} \E \| \wt{\bm{w}}_{k,i} \|^4 \le b_4^2 \cdot \nu^2
}
\ee
where the parameter $\nu$ is defined by 
\be
\label{eqn:nunewdef}
\nu \defeq \max_k \frac{\sqrt{ \bar{\mu}_k^{(4)} }}{ \bar{\mu}_k }, \quad b_4 \defeq \frac{ 3 \sigma_v^2 ( \kappa + 1)}{ \min_k \lambda_{k, \min} }
\ee
\end{theorem}
\begin{IEEEproof}
See Appendix \ref{app:4thorder}.
\end{IEEEproof}

It is easy to verify that condition \eqref{eqn:gradientnoise4thmoment} implies a bound on the second-order moment of the gradient noise:
\be
\label{eqn:gradientnoise2ndmoment}
\E[ \|\bm{v}_{k,i}(\bm{w}_{k,i-1}) \|^2 | \F_{i-1} ] \le \alpha  \| w^o - \bm{w}_{k,i-1}\|^2 + \sigma_v^2
\ee
although the converse is generally not true. In other words, it is redundant to assume both conditions \eqref{eqn:gradientnoise2norm} and \eqref{eqn:gradientnoise4thmoment}. It can be verified that condition \eqref{eqn:boundstepsize4thorder} implies \eqref{eqn:meansquarestabilitycond3} (see \eqref{eqn:oldnulessnewnu} and \eqref{eqn:newboundlessoldbound} in Appendix \ref{app:4thorder}). Therefore, conditions \eqref{eqn:gradientnoise4thmoment} and \eqref{eqn:boundstepsize4thorder} are sufficient to ensure both mean-square and fourth-order stability of error moments. Moreover, it is straightforward to verify that 
\be
\label{eqn:nuandnuo}
\nu_o = \max_k \frac{ \bar{\mu}_k^{(2)} }{ \bar{\mu}_k }
\le \max_k \frac{\sqrt{ \bar{\mu}_k^{(4)} }}{ \bar{\mu}_k }
= \nu
\ee
Therefore, we can use $\nu$ to upper bound $\nu_o$.

\section{Conclusion}
We introduced a fairly general model for \emph{asynchronous} behavior over networks with random step-sizes, links, topologies, and combination coefficients. We then carried out a mean-square analysis and showed that, even under non-vanishing step-sizes, the asynchronous network remains mean-square stable for sufficiently small step-sizes. We derived a condition on the first and second-order moments of the random step-sizes to ensure stable behavior. We specialized the results to two models: a Bernoulli network and a Beta network. It was observed that the Beta network admits a wider range of step-sizes for stability. The results suggest that networks where step-sizes assume values randomly within a certain interval are robustly more stable than networks that have their step-sizes be turned on or off. In Part II \cite{Zhao13TSPasync2} of this work, we derive explicit closed-form expressions for the MSD performance and use these expressions to clarify how the parameters of the asynchronous behavior influence both the convergence rate and the MSE performance. We will also be able to establish the useful clustering property illustrated by Fig. \ref{fig:normball}, namely, that the iterates at the various agents approach the optimal solution within accuracy $O(\nu)$ while clustering close to each other within $O(\nu^{1 + \gamma_o'})$ for some $\gamma_o' > 0$.

\appendices

\section{Equivalent Complex-Domain Representations}
\label{app:complex}
First, we recall the definition of the \emph{real} Jacobian of a real-valued function $J(w)$ with respect to a real column vector $w\in\mbbR^{M}$ as
\be
\label{eqn:realJacobiandef}
\frac{\partial J(w)}{\partial w} \defeq \row\left\{ \frac{\partial J(w)}{\partial w_1}, \frac{\partial J(w)}{\partial w_2}, \dots, \frac{\partial J(w)}{\partial w_M} \right\}
\ee
where $w_m\in\mbbR$ denotes the $m$-th element of $w$. Using \eqref{eqn:bardef}, the real gradient of the function $J_k(\bar{w})$ with respect to $\bar{w}$ is defined as
\be
\label{eqn:realgradientdef}
\nabla_{\bar{w}}J_k(\bar{w}) \defeq \frac{\partial J_k(\bar{w})}{\partial \bar{w}}
= \begin{bmatrix}
\frac{\partial J_k(w)}{\partial \Re(w)} & \frac{\partial J_k(w)}{\partial \Im(w)}
\end{bmatrix}
\ee
and the real Hessian matrix of the same function $J_k(\bar{w})$ with respect to $\bar{w}$ is defined by
\begin{align}
\label{eqn:realHessiandef}
\nabla_{\bar{w}\bar{w}^\T}^2J_k(\bar{w}) \defeq \frac{\partial}{\partial \bar{w}}\left[\frac{\partial J_k(\bar{w})}{\partial \bar{w}}\right]^\T = \begin{bmatrix}
\frac{\partial}{\partial \Re(w) }\left[\frac{\partial J_k(w)}{\partial \Re(w) }\right]^\T & 
\frac{\partial}{\partial \Im(w) }\left[\frac{\partial J_k(w)}{\partial \Re(w) }\right]^\T \\
\frac{\partial}{\partial \Re(w) }\left[\frac{\partial J_k(w)}{\partial \Im(w) }\right]^\T & 
\frac{\partial}{\partial \Im(w) }\left[\frac{\partial J_k(w)}{\partial \Im(w) }\right]^\T \\
\end{bmatrix}
\end{align}
It is easy to verify that $\nabla_{\bar{w}\bar{w}^\T}^2J_k(\bar{w})$ is a symmetric matrix.

Then, we define the derivative of a real-valued function $J(z)$ with respect to the complex argument $z\in\mbbC$ as \cite{Adali11TSP, vandenBos94VISP}:
\be
\label{eqn:complexderivative}
\frac{\partial J(z)}{\partial z} \defeq \frac{1}{2} \left( \frac{\partial J(z)}{\partial \Re(z)} - j \frac{\partial J(z)}{\partial \Im(z)} \right)
\ee
and the \emph{complex} Jacobian of a real-valued function $J(w)$ with respect to the complex column vector $w\in\mbbC^{M}$ is given by
\be
\label{eqn:complexJacobiandef}
\frac{\partial J(w)}{\partial w} \defeq \row\left\{ \frac{\partial J(w)}{\partial w_1}, \frac{\partial J(w)}{\partial w_2}, \dots, \frac{\partial J(w)}{\partial w_M} \right\}
\ee
where $w_m\in\mbbC$ denotes the $m$-th element of $w$. The complex gradient of the real-valued function $J_k(w)$ with respect to the complex vector argument $w\in\mbbC^{M}$ is defined as \cite[eq.~(20)]{vandenBos94VISP} (compare with \eqref{eqn:realgradientdef}):
\be
\label{eqn:complexgradientdef}
\nabla_w J_k(w) \defeq \frac{\partial J(w)}{\partial w} = \frac{1}{2} \left( \frac{\partial J(w)}{\partial \Re(w)} - j \frac{\partial J(w)}{\partial \Im(w)} \right)
\ee
and the complex conjugate gradient of $J_k(w)$ with respect to $w^*\in\mbbC^{M}$ is defined by \cite[eqs.~(21, 22)]{vandenBos94VISP}:
\be
\label{eqn:complexconjugategradientdef}
\nabla_{w^*} J_k(w) \defeq \frac{\partial J(w)}{\partial w^*} = \left[\nabla_w J_k(w)\right]^*
\ee
Using \eqref{eqn:ubardef}, the complex gradient of $J_k(\ubar{w})$ with respect to the column vector $\ubar{w}\in\mbbC_M^{2M}$ is then given by \cite[eq.~(18)]{vandenBos94VISP}:
\be
\label{eqn:complexgradientextdef}
\nabla_{\ubar{w}} J_k(\ubar{w}) = \frac{\partial J_k(\ubar{w})}{\partial \ubar{w}} = \begin{bmatrix}
\nabla_w J_k(w) & (\nabla_{w^*} J_k(w))^\T \\
\end{bmatrix}
\ee
and the corresponding complex conjugate gradient is given by
\be
\label{eqn:complexconjugategradientextdef}
\nabla_{\ubar{w}^*} J_k(\ubar{w}) = \left[ \frac{\partial J_k(\ubar{w})}{\partial \ubar{w}} \right]^* = [\nabla_{\ubar{w}} J_k(\ubar{w})]^*
\ee
The complex Hessian of $J_k(\ubar{w})$ with respect to $\ubar{w}\in\mbbC_M^{2M}$ is defined by \cite[eq.~(32)]{vandenBos94VISP}:
\begin{align}
\label{eqn:complexHessianextdef}
\nabla_{\ubar{w}\ubar{w}^{*}}^2 J_k(\ubar{w}) & \defeq \frac{\partial}{\partial \ubar{w}}\left[ \frac{\partial J_k(\ubar{w})}{\partial \ubar{w}} \right]^{*} \nn \\
& = \begin{bmatrix}
\nabla_{ww^*}^2 J_k(w) & (\nabla_{ww^\T}^2 J_k(w))^* \\
\nabla_{ww^\T}^2 J_k(w) & (\nabla_{ww^*}^2 J_k(w))^\T \\
\end{bmatrix}
\end{align}
where
\begin{align}
\nabla_{ww^*}^2 J_k(w) & \defeq \frac{\partial}{\partial w}\left[\frac{\partial J_k(w)}{\partial w}\right]^* = \begin{bmatrix}
\frac{\partial^2J_k(w)}{\partial w_1^*\partial w_1}  & \frac{\partial^2J_k(w)}{\partial w_1^*\partial w_2}  & \dots  & \frac{\partial^2J_k(w)}{\partial w_1^*\partial w_M} \\
\frac{\partial^2J_k(w)}{\partial w_2^*\partial w_1}  & \frac{\partial^2J_k(w)}{\partial w_2^*\partial w_2}  & \dots  & \frac{\partial^2J_k(w)}{\partial w_2^*\partial w_M} \\
\vdots  & \vdots  & \ddots  & \vdots \\
\frac{\partial^2J_k(w)}{\partial w_M^*\partial w_1}  & \frac{\partial^2J_k(w)}{\partial w_M^*\partial w_2}  & \dots  & \frac{\partial^2J_k(w)}{\partial w_M^*\partial w_M} \\
\end{bmatrix}\\
\nabla_{ww^\T}^2 J_k(w) & \defeq \frac{\partial}{\partial w}\left[\frac{\partial J_k(w)}{\partial w}\right]^\T = \begin{bmatrix}
\frac{\partial^2J_k(w)}{\partial w_1\partial w_1}  & \frac{\partial^2J_k(w)}{\partial w_1\partial w_2}  & \dots  & \frac{\partial^2J_k(w)}{\partial w_1\partial w_M} \\
\frac{\partial^2J_k(w)}{\partial w_2\partial w_1}  & \frac{\partial^2J_k(w)}{\partial w_2\partial w_2}  & \dots  & \frac{\partial^2J_k(w)}{\partial w_2\partial w_M} \\
\vdots  & \vdots  & \ddots  & \vdots \\
\frac{\partial^2J_k(w)}{\partial w_M\partial w_1}  & \frac{\partial^2J_k(w)}{\partial w_M\partial w_2}  & \dots  & \frac{\partial^2J_k(w)}{\partial w_M\partial w_M} \\
\end{bmatrix} 
\end{align}
It is easy to verify that $\nabla_{\ubar{w}\ubar{w}^{*}}^2 J_k(\ubar{w})$ is a Hermitian matrix.

From \eqref{eqn:realgradientdef} and \eqref{eqn:complexgradientextdef}, we have \cite[eqs.~(18, 19)]{vandenBos94VISP}:
\begin{align}
\label{eqn:realandcomplexGradient1}
\nabla_{\bar{w}}J_k(\bar{w}) & = \nabla_{\ubar{w}} J_k(\ubar{w}) \cdot D \\
\label{eqn:realandcomplexGradient2}
\nabla_{\bar{w}}J_k(\bar{w}) \cdot \frac{1}{2} D^* & = \nabla_{\ubar{w}} J_k(\ubar{w})
\end{align}
Similarly, from \eqref{eqn:realHessiandef} and \eqref{eqn:complexHessianextdef}, we have \cite[eqs.~(32, 33)]{vandenBos94VISP}:
\begin{align}
\label{eqn:realandcomplexHessian1}
\nabla_{\bar{w}\bar{w}^\T}^2J_k(\bar{w}) & = D^{*} \cdot [\nabla_{\ubar{w}\ubar{w}^{*}}^2 J_k(\ubar{w})] \cdot D \\
\label{eqn:realandcomplexHessian2}
\frac{1}{4}D \cdot [\nabla_{\bar{w}\bar{w}^\T}^2J_k(\bar{w})] \cdot D^{*} & = \nabla_{\ubar{w}\ubar{w}^{*}}^2 J_k(\ubar{w})
\end{align}
Identities \eqref{eqn:realandcomplexGradient1}--\eqref{eqn:realandcomplexHessian2} play an important role in our analysis.

\section{Proof of Lemma \ref{lemma:spatial}}
\label{app:spatial}
Expression \eqref{eqn:cov_mu_kl} is because $\bm{\mu}_k(i)$ and $\bm{\mu}_\ell(i)$ are uncorrelated when $k\ne \ell$. Expression \eqref{eqn:cov_mat_mu_kl} is obtained by using \eqref{eqn:cov_mu_kl} and \eqref{eqn:randstepsizecovblock}. Using \eqref{eqn:randomtopologyconstraints} and Lemma \ref{lemma:neighborhoods}, we have
\be
\label{eqn:akkidef}
\bm{a}_{k k}(i) = 1 - \sum_{\ell \in \bm{\Ncal}_{k,i}\backslash\{k\}} \bm{a}_{\ell k}(i) = 1 - \sum_{\ell \in \Nknk} \bm{a}_{\ell k}(i)
\ee
since $\bm{a}_{\ell k}(i) = 0$ for any $\ell \in \Ncal_k\backslash\bm{\Ncal}_{k,i}$. When $\ell \ne k$, all entries in $\bm{A}_i$ are uncorrelated with $\bm{a}_{\ell k}(i)$ except for the $(\ell,k)$-th and $(k,k)$-th entries. It follows from Lemma \ref{lemma:neighborhoods} that
\begin{align}
\label{eqn:uncorrelated_A_cov_lk_kk}
c_{a,\ell k, kk} & = \E[(\bm{a}_{\ell k}(i) - \bar{a}_{\ell k})(\bm{a}_{k k}(i) - \bar{a}_{k k}) ] \nn \\
& \stackrel{(a)}{=} -\sum_{n\in\Nknk} \E[(\bm{a}_{\ell k}(i) - \bar{a}_{\ell k})(\bm{a}_{nk}(i) - \bar{a}_{nk}) ] \nn \\
& \stackrel{(b)}{=} - \, \E(\bm{a}_{\ell k}(i) - \bar{a}_{\ell k})^2 \nn \\
& \stackrel{(c)}{=} - \, c_{a,\ell k, \ell k}
\end{align}
for any $\ell \in \Nknk$ since \eqref{eqn:uncorrelated_A_cov_lk_kk} holds for any realization of the random neighborhood $\bm{\Ncal}_{k,i}$, and where step (a) is due to \eqref{eqn:akkidef}; step (b) is because $\{\bm{a}_{nk}(i); n \in \bm{\Ncal}_{k,i}\backslash\{k\}\}$ are all uncorrelated with $\bm{a}_{\ell k}(i)$ except for $\bm{a}_{\ell k}(i)$ itself, and $\bm{a}_{\ell k}(i) = 0$ for any $\ell \in \Ncal_k \backslash \bm{\Ncal}_{k,i}$; and step (c) is because of \eqref{eqn:randcombinecoventry}. From \eqref{eqn:uncorrelated_A_cov_lk_kk}, we get \eqref{eqn:cov_a_lk_nm}. When $\ell = k$, all entries in $\bm{A}_i$ are uncorrelated with $\bm{a}_{kk}(i)$ except for the $(\ell,k)$-th entries for all $\ell \in \bm{\Ncal}_{k,i}$. It follows from Lemma \ref{lemma:neighborhoods} that
\begin{align}
\label{eqn:uncorrelated_A_cov_kk_lk}
c_{a,kk, \ell k} & = \E[(\bm{a}_{kk}(i) - \bar{a}_{kk})(\bm{a}_{\ell k}(i) - \bar{a}_{\ell k}) ] \nn \\
& = - \, c_{a,\ell k, \ell k}, \qquad \ell\in\Nknk \\
\label{eqn:uncorrelated_A_cov_kk_kk}
c_{a,kk, kk} & \stackrel{(a)}{=} \sum_{\ell,n\in\Nknk} \E [ (\bm{a}_{\ell k}(i) - \bar{a}_{\ell k})(\bm{a}_{nk}(i) - \bar{a}_{nk}) ] \nn \\
& \stackrel{(b)}{=} \sum_{\ell\in\Nknk} \E (\bm{a}_{\ell k}(i) - \bar{a}_{\ell k})^2 \nn \\
& \stackrel{(c)}{=} \sum_{\ell\in\Nknk} c_{a,\ell k, \ell k} 
\end{align}
where \eqref{eqn:uncorrelated_A_cov_kk_lk} is because of \eqref{eqn:uncorrelated_A_cov_lk_kk}; step (a) is because of \eqref{eqn:akkidef}; step (b) is because $\{\bm{a}_{\ell k}(i); \ell \in \bm{\Ncal}_{k,i}\backslash\{k\}\}$ are mutually-uncorrelated, and $\bm{a}_{\ell k}(i) = 0$ for any $\ell \in \Ncal_k \backslash \bm{\Ncal}_{k,i}$; and step (c) is because of \eqref{eqn:randcombinecoventry}. From \eqref{eqn:uncorrelated_A_cov_kk_lk} and \eqref{eqn:uncorrelated_A_cov_kk_kk}, we get \eqref{eqn:cov_a_lk_nm}.

\section{Derivation of Error Recursion \eqref{eqn:atcasync1errorubar1}--\eqref{eqn:atcasync2errorubar1}}
\label{app:errorrecursion}
Applying the transformation $\ubar\mbbT$ from \eqref{eqn:ubardef} to both sides of the error recursion \eqref{eqn:atcasync1error}--\eqref{eqn:atcasync2error}, we get
\begin{subequations}
\begin{align}
\label{eqn:atcasync1errorubar}
\ubar{\wt{\bm{\psi}}}_{k,i} &  =  \ubar{\wt{\bm{w}}}_{k,i - 1}  +  \bm{\mu}_k(i)[\nabla_{\ubar{w}^*}J_k(\ubar{\bm{w}}_{k,i - 1})  +  \ubar{\bm{v}}_{k,i}(\bm{w}_{k,i - 1}) ] \\
\label{eqn:atcasync2errorubar}
\ubar{\wt{\bm{w}}}_{k,i} &  =  \sum_{\ell\in\bm{\Ncal}_{k,i}} \bm{a}_{\ell k}(i)\,\ubar{\wt{\bm{\psi}}}_{\ell,i}
\end{align}
\end{subequations}
where, by definition,
\be
\ubar{\mbbT}(\nabla_{w^*}J_k(w)) = \begin{bmatrix}
\nabla_{w^*}J_k(w) \\
\nabla_{w^\T}J_k(w)
\end{bmatrix} = \nabla_{\ubar{w}^*}J_k(\ubar{w})
\ee
The \emph{real} gradient defined by \eqref{eqn:realgradientdef} can be expressed using the mean-value theorem as \cite{Polyak87}:
\be
\label{eqn:realgradientmidvalue}
\nabla_{\bar{w}^\T}J_k(\bar{w}) = \left[\int_{0}^{1}\nabla_{\bar{w}\bar{w}^\T}^2 J_k(\bar{w}^o - t(\bar{w}^o - {\bar{w}}))dt\right]  (\bar{w} - \bar{w}^o)
\ee
since $\nabla_{\bar{w}^\T}J_k(\bar{w}^o) = 0$ by Assumption \ref{asm:costfunctions}. From \eqref{eqn:barw2ubarw}, \eqref{eqn:realgradientmidvalue}, \eqref{eqn:realandcomplexGradient2}, and \eqref{eqn:realandcomplexHessian2}, we get
\begin{align}
\nabla_{\ubar{w}^*}J_k(\ubar{w})
& = \frac{1}{2}D \cdot \nabla_{\bar{w}^\T}J_k(\bar{w}) \nn \\
& = \int_{0}^{1} \frac{1}{4} D \left[\nabla_{\bar{w}\bar{w}^\T}^2J_k(\bar{w}^o-t(\bar{w}^o-{\bar{w}}))\right] D^{*} dt \cdot D (\bar{w}-\bar{w}^o) \nn \\
& = \left[\int_{0}^{1} \nabla_{\ubar{w}\ubar{w}^*}^2J_k(\ubar{w}^o-t(\ubar{w}^o-{\ubar{w}})) dt \right] \cdot (\ubar{w}-\ubar{w}^o)
\end{align}
Letting $\ubar{w} = \ubar{\bm{w}}_{k,i-1}$, we get
\be
\label{eqn:gradientandhessian}
\nabla_{\ubar{w}^*}J_k(\ubar{\bm{w}}_{k,i-1}) = -   \left[\int_{0}^{1}\nabla_{\ubar{w}\ubar{w}^*}^2 J_k(\ubar{w}^o  -  t\ubar{\wt{\bm{w}}}_{k,i-1})\,dt \right] \ubar{\wt{\bm{w}}}_{k,i-1}
\ee
Then, by \eqref{eqn:gradientandhessian}, the error recursion \eqref{eqn:atcasync1errorubar} and \eqref{eqn:atcasync2errorubar} can be rewritten as \eqref{eqn:atcasync1errorubar1}--\eqref{eqn:atcasync2errorubar1}.

\section{Proof of Theorem \ref{theorem:meansquarestability}}
\label{app:proofofstablity}
We start from equation \eqref{eqn:atcasync2errorubar1}. Since the squared Euclidean norm $\|\cdot\|^2$ is a convex function of its vector argument, using Jensen's inequality \cite{Boyd04} we get
\be
\label{eqn:meansquareboundw0}
\|\ubar{\wt{\bm{w}}}_{k,i}\|^2 \le \sum_{\ell\in \bm{\Ncal}_{k,i}} \bm{a}_{\ell k}(i)\|\ubar{\wt{\bm{\psi}}}_{\ell,i} \|^2 = \sum_{\ell\in \Ncal_k} \bm{a}_{\ell k}(i)\|\ubar{\wt{\bm{\psi}}}_{\ell,i} \|^2   
\ee
since $\bm{a}_{\ell k}(i) = 0$ for any $\ell \in \Ncal_k\backslash\bm{\Ncal}_{k,i}$ by \eqref{eqn:randomtopologyconstraints} and Lemma \ref{lemma:neighborhoods}. Taking the expectation of both sides of \eqref{eqn:meansquareboundw0} and using the asynchronous network model, we get
\be
\label{eqn:meansquareboundw2}
\E\|\ubar{\wt{\bm{w}}}_{k,i}\|^2 \le \sum_{\ell\in{\Ncal}_k} \bar{a}_{\ell k} \, \E\|\ubar{\wt{\bm{\psi}}}_{\ell,i} \|^2  \le \max_\ell \{ \E\|\ubar{\wt{\bm{\psi}}}_{\ell,i} \|^2 \}
\ee
Conditioned on $\F_{i-1}$, the random matrix $\bm{H}_{k,i-1}$ defined by \eqref{eqn:Hkidef} becomes deterministic. Let
\be
\label{eqn:Sigmakidef}
\bm{\Sigma}_{k,i} \defeq [I_{2M}-\bm{\mu}_k(i)\bm{H}_{k,i-1}]^2
\ee
From \eqref{eqn:atcasync1errorubar1}, we get
\begin{align}
\label{eqn:meansquareboundpsi1}
\E(\|\ubar{\wt{\bm{\psi}}}_{\ell,i} \|^2 | \F_{i-1} )
& \stackrel{(a)}{=} \E(\|\ubar{\wt{\bm{w}}}_{k,i-1}\|_{\bm{\Sigma}_{k,i}}^2 | \F_{i-1} ) + \E[\bm{\mu}_k^2(i) \|\ubar{\bm{v}}_{k,i}(\bm{w}_{k,i-1})\|^2 | \F_{i-1} ] \nn \\
& \stackrel{(b)}{\le} \E(\|\bm{\Sigma}_{k,i}\| \cdot \|\ubar{\wt{\bm{w}}}_{k,i-1}\|^2 | \F_{i-1} ) + (\bar{\mu}_k^2 + c_{\mu,k,k})\cdot\E[\|\ubar{\bm{v}}_{k,i}(\bm{w}_{k,i-1})\|^2 | \F_{i-1} ] \nn \\
& \stackrel{(c)}{\le} \E(\|\bm{\Sigma}_{k,i}\| | \F_{i-1} ) \cdot \|\ubar{\wt{\bm{w}}}_{k,i-1}\|^2  + (\bar{\mu}_k^2 + c_{\mu,k,k}) \cdot ( \alpha \cdot \|\ubar{\wt{\bm{w}}}_{k,i-1}\|^2 + 2\sigma_v^2 )
\end{align}
where step (a) is from \eqref{eqn:Sigmakidef} and cross terms are eliminated by using the conditional independence and zero-mean properties of $\bm{v}_{k,i}(\bm{w}_{k,i-1})$ from Assumption \ref{asm:gradientnoise}; step (b) in \eqref{eqn:meansquareboundpsi1} is due to the asynchronous network model and the sub-multiplicative property of the 2-norm; and step (c) is by conditioning and \eqref{eqn:gradientnoise2norm}. Using Assumptions \ref{asm:boundedHessian} and \eqref{eqn:Hkidef} we have
\be
1  -  \bm{\mu}_k(i)\lambda_{k,\max} \le \lambda( I_{2M}  -  \bm{\mu}_k(i) \bm{H}_{k,i-1} ) \le 1  -  \bm{\mu}_k(i)\lambda_{k,\min}
\ee
Then, from \eqref{eqn:Sigmakidef}, we obtain
\begin{align}
\label{eqn:boundeiglambda}
\lambda\left( \bm{\Sigma}_{k,i} \right) & \le \max\{ (1-\bm{\mu}_k(i)\lambda_{k,\min})^2, (1-\bm{\mu}_k(i)\lambda_{k,\max})^2 \} \nn \\
& = \max\{ 1 - 2\bm{\mu}_k(i)\lambda_{k,\min} + \bm{\mu}_k^2(i) \lambda_{k,\min}^2, 1 - 2\bm{\mu}_k(i)\lambda_{k,\max} + \bm{\mu}_k^2(i)\lambda_{k,\max}^2 \} \nn \\
& \le 1 - 2\bm{\mu}_k(i)\lambda_{k,\min} + \bm{\mu}_k^2(i)\lambda_{k,\max}^2
\end{align}
because $\bm{\mu}_k(i)$ is nonnegative. Therefore, we have
\begin{align}
\label{eqn:Sigmakinormbound}
\E(\|\bm{\Sigma}_{k,i}\| | \F_{i-1} ) & \stackrel{(a)}{=} \E[ \lambda_{\max}\left( \bm{\Sigma}_{k,i} \right) | \F_{i-1} ]  \nn \\
& \stackrel{(b)}{\le} \E[1 - 2\bm{\mu}_k(i)\lambda_{k,\min} + \bm{\mu}_k^2(i)\lambda_{k,\max}^2] \nn \\
& \stackrel{(c)}{=} \gamma_k^2
\end{align}
where step (a) is because $\bm{\Sigma}_{k,i}$ in \eqref{eqn:Sigmakidef} is Hermitian and positive semi-definite, and its largest singular value coincides with its largest eigenvalue; step (b) is by using \eqref{eqn:boundeiglambda} and the independence condition in the asynchronous model; and step (c) is by \eqref{eqn:gammak2def}. Substituting \eqref{eqn:Sigmakinormbound} into \eqref{eqn:meansquareboundpsi1}, and taking the expectation of both sides with respect to $\ubar{\wt{\bm{w}}}_{i-1}$ yields
\begin{align}
\label{eqn:meansquareboundpsi2}
\E \| \ubar{\wt{\bm{\psi}}}_{k,i} \|^2 & \le [\gamma_k^2 + \alpha(\bar{\mu}_k^2 + c_{\mu,k,k} )] \cdot \E\|\ubar{\wt{\bm{w}}}_{k,i-1}\|^2  + 2 (\bar{\mu}_k^2 + c_{\mu,k,k} ) \sigma_v^2
\end{align}
Combining \eqref{eqn:meansquareboundpsi2} and \eqref{eqn:meansquareboundw2} yields
\begin{align}
\label{eqn:meansquareerrorboundext}
\E\|\ubar{\wt{\bm{w}}}_{k,i}\|^2 & \le \max_\ell\{[\gamma_\ell^2 + \alpha(\bar{\mu}_\ell^2 + c_{\mu,\ell,\ell})] \cdot \E\|\ubar{\wt{\bm{w}}}_{\ell,i-1}\|^2  + 2(\bar{\mu}_\ell^2 + c_{\mu,\ell,\ell} ) \sigma_v^2 \}
\end{align}
Dividing both sides of \eqref{eqn:meansquareerrorboundext} by 2 and using the fact that $\E\,\|\wt{\bm{w}}_{k,i-1}\|^2 = \E\|\ubar{\wt{\bm{w}}}_{k,i-1}\|^2 / 2$, we get
\begin{align}
\label{eqn:meansquareerrorbound0}
\E\|{\wt{\bm{w}}}_{k,i}\|^2 & \le \left[\max_\ell\{\gamma_\ell^2 + \alpha(\bar{\mu}_\ell^2 + c_{\mu,\ell,\ell} )\}\right] \left[\max_\ell \E\| \wt{\bm{w}}_{\ell,i-1} \|^2 \right] + \left[\max_\ell\{\bar{\mu}_\ell^2 + c_{\mu,\ell,\ell} \}\right] \cdot \sigma_v^2 
\end{align}
Now since inequality \eqref{eqn:meansquareerrorbound0} holds for every $k$, using \eqref{eqn:epsilondef}, we conclude that \eqref{eqn:meansquareerrorboundrecursion} should hold. Propagating \eqref{eqn:meansquareerrorboundrecursion} backwards to the starting point yields
\be
\label{eqn:meansquareerrorbound2}
\epsilon^2(i) \le \beta^{i+1} \cdot \epsilon^2(-1) + \theta\sigma_v^2\cdot\sum_{j=0}^{i}\beta^j
\ee
where $\epsilon^2(-1) \defeq \max_k \E\,\|\wt{\bm{w}}_{k,-1}\|^2$ represents the initial error variance. In order to guarantee a convergent upper bound, we require $|\beta| < 1$, which, by \eqref{eqn:gammak2def} and \eqref{eqn:betadef}, is equivalent to
\be
\label{eqn:meansquarestabilitycond2}
|1 - 2\bar{\mu}_k\lambda_{k,\min} +  ( \bar{\mu}_k^2 + c_{\mu, k, k})( \lambda_{k,\max}^2 + \alpha )| < 1 
\ee
for any $k$. A sufficient condition for  \eqref{eqn:meansquarestabilitycond2} is given by
\be
\label{eqn:meansquareerrorbound2new}
\frac{\bar{\mu}_k^2 + c_{\mu,k,k}}{\bar{\mu}_k} < \frac{2\lambda_{k,\min}}{\alpha + \lambda_{k,\max}^2}
\ee
It is easy to verify that condition \eqref{eqn:meansquarestabilitycond3} is a sufficient condition for  \eqref{eqn:meansquareerrorbound2new}. Therefore, if condition \eqref{eqn:meansquarestabilitycond3} holds, then $|\beta| < 1$.

Now under condition \eqref{eqn:meansquareerrorbound2new}, we obtain from  \eqref{eqn:meansquareerrorboundrecursion} that
\begin{align}
\label{eqn:meansquareerrorbound3}
\epsilon^2(i) \le \beta^{i+1} \cdot \epsilon^2(-1) + \frac{\theta\sigma_v^2(1-\beta^{i+1})}{1-\beta}
\end{align}
When $i\rightarrow\infty$, we get an upper bound for the individual MSD:
\be
\label{eqn:individualMSDupboundnew1}
\limsup_{i\rightarrow\infty} \epsilon^2(i) \le \frac{\theta\sigma_v^2}{1-\beta}
\ee
In the following we simplify the upper bound in \eqref{eqn:individualMSDupboundnew1}. From \eqref{eqn:betadef} and \eqref{eqn:gammak2def}, we get
\begin{align}
\label{eqn:1minusbeta}
1 - \beta 
& = 1 - \max_k\{ \gamma_k^2 + \alpha ( \bar{\mu}_k^2 + c_{\mu, k, k}) \} \nn \\
& = 1 - \max_k\{ 1 - 2 \bar{\mu}_k \lambda_{k,\min} + ( \bar{\mu}_k^2 + c_{\mu, k, k}) ( \lambda_{k,\max}^2 + \alpha ) \} \nn \\
& = \min_k \left\{ \bar{\mu}_k \cdot \left[2\lambda_{k,\min} - \frac{\bar{\mu}_k^2 + c_{\mu,k,k}}{\bar{\mu}_k} (\alpha + \lambda_{k,\max}^2) \right] \right\} \nn \\
& \ge \min_k\{ \bar{\mu}_k \} \cdot \min_k \left[2\lambda_{k,\min} - \frac{\bar{\mu}_k^2 + c_{\mu,k,k}}{\bar{\mu}_k} (\alpha + \lambda_{k,\max}^2) \right]
\end{align}
Using \eqref{eqn:meansquarestabilitycond3} again, we get
\be
\frac{\bar{\mu}_k^2 + c_{\mu,k,k}}{\bar{\mu}_k} (\alpha + \lambda_{k,\max}^2) < \lambda_{k,\min}
\ee
Hence, relation \eqref{eqn:1minusbeta} can be further expressed as
\be
\label{eqn:1minusbeta1}
1 - \beta \ge \min_k\{ \bar{\mu}_k \} \cdot \min_k\{ \lambda_{k,\min} \}
\ee
From \eqref{eqn:thetadef} we get
\be
\label{eqn:thetabound}
\theta \le \max_k \frac{\bar{\mu}_k^2 + c_{\mu,k,k}}{\bar{\mu}_k} \cdot \max_k \bar{\mu}_k 
\ee
Therefore, when $i\rightarrow\infty$, using \eqref{eqn:1minusbeta1}, \eqref{eqn:thetabound}, \eqref{eqn:kappadef}, and \eqref{eqn:mumaxdef}, we get from \eqref{eqn:individualMSDupboundnew1} that
\begin{align}
\label{eqn:meansquareerrorbound4}
\limsup_{i\rightarrow\infty}\epsilon^2(i) & \le \frac{\theta\sigma_v^2}{1-\beta} \nn \\
& \le \frac{\sigma_v^2}{\min_k\{\lambda_{k,\min}\}} \frac{\max_k\{\bar{\mu}_k\}}{\min_k\{\bar{\mu}_k\}} \max_k \frac{\bar{\mu}_k^2 + c_{\mu,k,k}}{\bar{\mu}_k} \nn \\
& \le \frac{\kappa \sigma_v^2}{\min_k\{\lambda_{k,\min}\}} \cdot \max_k \frac{\bar{\mu}_k^2 + c_{\mu,k,k}}{\bar{\mu}_k}
\end{align}
Substituting \eqref{eqn:mumaxdef} into \eqref{eqn:meansquareerrorbound4} completes the proof. 

\section{Proof of Theorem \ref{theorem:4thmoments}}
\label{app:4thorder}
From \eqref{eqn:atcasync2errorubar1} and using Jensen's inequality, we obtain under expectation:
\be
\label{eqn:bound4thorderequ1}
\E \| \ubar{\wt{\bm{w}}}_{k,i} \|^4 \le \sum_{\ell\in\Ncal_k} \bar{a}_{\ell k}\,\E\| \ubar{\wt{\bm{\psi}}}_{\ell,i} \|^4
\ee
for all $k$. Therefore, we have
\begin{align}
\label{eqn:bound4thorderequmax}
\max_k \E \| \ubar{\wt{\bm{w}}}_{k,i} \|^4 & \le \max_k \E\| \ubar{\wt{\bm{\psi}}}_{k,i} \|^4
\end{align}
From \eqref{eqn:atcasync1errorubar1}, we have
\begin{align}
\label{eqn:exppsi4thorder}
\| \ubar{\wt{\bm{\psi}}}_{k,i} \|^4 & = \| [I_{2M} - \bm{\mu}_k(i)\bm{H}_{k,i - 1}]\ubar{\wt{\bm{w}}}_{k,i-1}  + \bm{\mu}_k(i)\ubar{\bm{v}}_{k,i}(\bm{w}_{k,i-1}) \|^4
\end{align}

\begin{lemma}[Fourth-order inequality]
\label{lemma:quarticnorm}
For any two vectors $\bm{x}$ and $\bm{y}$ of the same size, it holds that
\be
\label{eqn:bound4thnorm}
\| \bm{x} + \bm{y} \|^4 \le \| \bm{x} \|^4 + 8 \| \bm{x} \|^2 \| \bm{y} \|^2 + 3\| \bm{y} \|^4 + 4 \| \bm{x} \|^2 \Re(\bm{x}^*\bm{y})
\ee
\end{lemma}
\begin{IEEEproof}
It holds that
\begin{align}
\| \bm{x} + \bm{y} \|^4 & = [ \| \bm{x} \|^2 + 2\Re(\bm{x}^*\bm{y}) + \| \bm{y} \|^2]^2 \nn \\
& = \| \bm{x} \|^4 + 4[\Re(\bm{x}^*\bm{y})]^2 + \| \bm{y} \|^4 + 2 \| \bm{x} \|^2 \| \bm{y} \|^2  + 4 \| \bm{x} \|^2 \Re(\bm{x}^*\bm{y}) + 4 \Re(\bm{x}^*\bm{y}) \| \bm{y} \|^2
\end{align}
The result now follows by using the inequalities:
\be
|\Re(\bm{x}^*\bm{y})|^2 \le \| \bm{x} \|^2 \| \bm{y} \|^2, \;\;
2\Re(\bm{x}^*\bm{y}) \le \| \bm{x} \|^2 + \| \bm{y} \|^2
\ee
\end{IEEEproof}

Referring to \eqref{eqn:exppsi4thorder}, if we make the identifications
\be
\label{eqn:xandydef}
\bm{x} \equiv [I_{2M} - \bm{\mu}_k(i)\bm{H}_{k,i - 1}]\ubar{\wt{\bm{w}}}_{k,i-1}, \;\; \bm{y} \equiv  \bm{\mu}_k(i)\ubar{\bm{v}}_{k,i}(\bm{w}_{k,i-1})
\ee
then we obtain
\begin{align}
\label{eqn:x4thorder}
\| \bm{x} \|^2 \le \bm{a} \cdot \bm{b}, \qquad \| \bm{y} \|^2 = \bm{c} \cdot \bm{d}
\end{align}
where
\begin{align}
\bm{a} & \defeq 1 - 2\bm{\mu}_k(i) \lambda_{k, \min} + \bm{\mu}_k^2(i) \lambda_{k, \max}^2 \\
\label{eqn:bdef}
\bm{b} & \defeq \| \ubar{\wt{\bm{w}}}_{k,i-1} \|^2 \\
\bm{c} & \defeq \bm{\mu}_k^2(i) \\
\bm{d} & \defeq \|\ubar{\bm{v}}_{k,i}(\bm{w}_{k,i-1}) \|^2
\end{align}
Using Lemma \ref{lemma:quarticnorm}, we obtain from \eqref{eqn:x4thorder} that
\be
\label{eqn:bound4thorder1}
\| \bm{x} + \bm{y} \|^4 \le \bm{a}^2 \cdot \bm{b}^2 + 8 \bm{a} \cdot \bm{b} \cdot \bm{c} \cdot \bm{d} + 3 \bm{c}^2 \cdot \bm{d}^2 + 4 \| \bm{x} \|^2 \Re(\bm{x}^*\bm{y})
\ee
where
\begin{align}
\label{eqn:a2expression}
\bm{a}^2 & = [1 - 2\bm{\mu}_k(i) \lambda_{k, \min} + \bm{\mu}_k^2(i) \lambda_{k, \max}^2]^2 \nn \\
& = 1 - 4\bm{\mu}_k(i) \lambda_{k, \min} + 2\bm{\mu}_k^2(i)(2\lambda_{k, \min}^2 + \lambda_{k, \max}^2) - 4\bm{\mu}_k^3(i) \lambda_{k, \min}\lambda_{k, \max}^2 + \bm{\mu}_k^4(i) \lambda_{k, \max}^4 \nn \\
& < 1 - 4\bm{\mu}_k(i) \lambda_{k, \min} + 2\bm{\mu}_k^2(i)(2\lambda_{k, \min}^2 + \lambda_{k, \max}^2)  + \bm{\mu}_k^4(i) \lambda_{k, \max}^4 \\
\label{eqn:c2expression}
\bm{c}^2 & = \bm{\mu}_k^4(i) \\
\label{eqn:acexpression}
\bm{a} \cdot \bm{c} & = \bm{\mu}_k^2(i) - 2\bm{\mu}_k^3(i) \lambda_{k, \min} + \bm{\mu}_k^4(i) \lambda_{k, \max}^2 \nn \\
& \le \bm{\mu}_k^2(i) + \bm{\mu}_k^4(i) \lambda_{k, \max}^2
\end{align}
Taking the expectation of \eqref{eqn:bound4thorder1} conditioned on $\F_{i-1}$, we get
\be
\label{eqn:exppsi4thorder1}
\E [\| \bm{x} + \bm{y} \|^4 | \F_{i-1}] \le \E[\bm{a}^2] \cdot \bm{b}^2 + 8 \E[\bm{a} \cdot \bm{c}] \cdot \bm{b} \cdot \E[ \bm{d}] + 3 \E[\bm{c}^2] \cdot \E[\bm{d}^2]
\ee
where the last term disappears because $\bm{y}$ has the noise factor that is conditionally zero mean. From \eqref{eqn:a2expression}--\eqref{eqn:acexpression}, we have
\begin{align}
\label{eqn:bounda2}
\E[\bm{a}^2] & \le 1 - 4\bar{\mu}_k^{(1)} \lambda_{k, \min} + 2 \bar{\mu}_k^{(2)} (2\lambda_{k, \min}^2 + \lambda_{k, \max}^2)  + \bar{\mu}_k^{(4)} \lambda_{k, \max}^4 \\
\label{eqn:boundc2}
\E[\bm{c}^2] & = \bar{\mu}_k^{(4)} \\
\label{eqn:boundac}
\E[\bm{a} \cdot \bm{c}] & \le \bar{\mu}_k^{(2)} + \bar{\mu}_k^{(4)} \lambda_{k, \max}^2
\end{align}
where $\bar{\mu}_k^{(m)} \defeq \E[\bm{\mu}_k^m(i)]$ denotes the $m$-th moment of the random step-size parameter $\bm{\mu}_k(i)$. It follows from \eqref{eqn:gradientnoise4thmoment} that
\be
\label{eqn:bound4thnormnoise}
\E[ \| \ubar{\bm{v}}_{k,i}(\bm{w}_{k,i-1}) \|^4 | \F_{i-1} ] \le \alpha^2 \cdot \| \ubar{\wt{\bm{w}}}_{k,i-1}\|^4 + 4 \sigma_v^4
\ee
where a factor of $4$ appears because of the transform $\ubar{\mbbT}(\cdot)$. Likewise, it follows from \eqref{eqn:gradientnoise2ndmoment} that
\be
\label{eqn:bound2ndnormnoise}
\E[ \| \ubar{\bm{v}}_{k,i}(\bm{w}_{k,i-1}) \|^2 | \F_{i-1} ] \le \alpha \cdot \| \ubar{\wt{\bm{w}}}_{k,i-1}\|^2 + 2\sigma_v^2
\ee 
Using \eqref{eqn:bound4thnormnoise} and \eqref{eqn:bound2ndnormnoise}, we can bound the quantities $\E[\bm{d}^2]$ and $\E[\bm{d}]$ in \eqref{eqn:exppsi4thorder1} by
\begin{align}
\label{eqn:boundd2}
\E[\bm{d}^2] & \le \alpha^2 \cdot \| \ubar{\wt{\bm{w}}}_{k,i-1}\|^4 + 4\sigma_v^4 = \alpha^2 \cdot \bm{b}^2 + 4\sigma_v^4\\
\label{eqn:boundd}
\E[\bm{d}] & \le \alpha \cdot \| \ubar{\wt{\bm{w}}}_{k,i-1}\|^2 + 2\sigma_v^2 = \alpha \cdot \bm{b} + 2\sigma_v^2
\end{align}
Substituting \eqref{eqn:bounda2}--\eqref{eqn:boundac} and \eqref{eqn:boundd2}--\eqref{eqn:boundd} into \eqref{eqn:exppsi4thorder1}, we end up with
\begin{align}
\label{eqn:exppsi4thorder2}
\E [\| \bm{x} + \bm{y} \|^4 | \F_{i-1}] 
& \le [1 - 4\bar{\mu}_k^{(1)} \lambda_{k, \min} + 2 \bar{\mu}_k^{(2)} (2\lambda_{k, \min}^2 + \lambda_{k, \max}^2) + \bar{\mu}_k^{(4)} \lambda_{k, \max}^4] \bm{b}^2 \nn \\
& \qquad + 8 [\bar{\mu}_k^{(2)} + \bar{\mu}_k^{(4)} \lambda_{k, \max}^2] \cdot \bm{b} \cdot (\alpha \cdot \bm{b} + 2\sigma_v^2)  + 3 \bar{\mu}_k^{(4)} \cdot (\alpha^2 \cdot \bm{b}^2 + 4 \sigma_v^4) \nn \\
& = [1 - 4\bar{\mu}_k^{(1)} \lambda_{k, \min} + 2 \bar{\mu}_k^{(2)} (2\lambda_{k, \min}^2 + \lambda_{k, \max}^2 + 4\alpha) \nn \\
& \qquad + \bar{\mu}_k^{(4)} (\lambda_{k, \max}^4 + 8\alpha \lambda_{k, \max}^2 + 3 \alpha^2)] \cdot \bm{b}^2 + 16 \sigma_v^2 [\bar{\mu}_k^{(2)} + \bar{\mu}_k^{(4)} \lambda_{k, \max}^2] \cdot \bm{b}  + 12 \sigma_v^4 \bar{\mu}_k^{(4)} \nn \\
& \defeq (1 - h_{k,1}) \cdot \bm{b}^2 + h_{k,2} \cdot \bm{b} + h_{k,3}
\end{align}
where
\begin{align}
\label{eqn:hk1def}
h_{k,1} & \defeq 4\bar{\mu}_k^{(1)} \lambda_{k, \min} - 2 \bar{\mu}_k^{(2)} (2\lambda_{k, \min}^2 + \lambda_{k, \max}^2 + 4\alpha) - \bar{\mu}_k^{(4)} (\lambda_{k, \max}^4 + 8\alpha \lambda_{k, \max}^2 + 3 \alpha^2) \\
\label{eqn:hk2def}
h_{k,2} & \defeq 16 \sigma_v^2 (\bar{\mu}_k^{(2)} + \bar{\mu}_k^{(4)} \lambda_{k, \max}^2) \\
\label{eqn:hk3def}
h_{k,3} & \defeq 12 \sigma_v^4 \bar{\mu}_k^{(4)}
\end{align}
Substituting \eqref{eqn:xandydef}, \eqref{eqn:exppsi4thorder}, and \eqref{eqn:bdef} into \eqref{eqn:exppsi4thorder2}, we get
\begin{align}
\label{eqn:exppsi4thorder3}
\E [\| \ubar{\wt{\bm{\psi}}}_{k,i} \|^4 | \F_{i-1}] & \le (1 - h_{k,1}) \cdot \| \ubar{\wt{\bm{w}}}_{k,i-1} \|^4 + h_{k,2} \cdot \| \ubar{\wt{\bm{w}}}_{k,i-1} \|^2 + h_{k,3}
\end{align}
Taking the expectation with respect to $\F_{i-1}$ yields
\begin{align}
\label{eqn:exppsi4thorder4}
\E\| \ubar{\wt{\bm{\psi}}}_{k,i} \|^4 & \le (1 - h_{k,1}) \cdot \E \| \ubar{\wt{\bm{w}}}_{k,i-1} \|^4 + h_{k,2} \cdot \E \| \ubar{\wt{\bm{w}}}_{k,i-1} \|^2 + h_{k,3}
\end{align}
From \eqref{eqn:meansquareerrorbound7} in Theorem \ref{theorem:meansquarestability}, we know for large enough $i$ that
\be
\E\|\ubar{\wt{\bm{w}}}_{k,i-1}\|^2 \le  2(b  + \epsilon) \cdot \nu
\ee
where we used the fact that $\|\ubar{w}\|^2 = 2\|w\|^2$, and $0 < \epsilon \ll 1$ is a small number. Therefore, we can bound $\E\| \ubar{\wt{\bm{\psi}}}_{k,i} \|^4$ in \eqref{eqn:exppsi4thorder4} for large enough $i$ by
\begin{align}
\label{eqn:exppsi4thorder5}
\E\| \ubar{\wt{\bm{\psi}}}_{k,i} \|^4 & \le (1 - h_{k,1}) \cdot \E \| \ubar{\wt{\bm{w}}}_{k,i-1} \|^4 + h_{k,2} \cdot 2(b  + \epsilon) \cdot \nu + h_{k,3}
\end{align}
Substituting \eqref{eqn:exppsi4thorder5} into \eqref{eqn:bound4thorderequmax}, we get
\begin{align}
\label{eqn:bound4thorderequmax1}
\max_k \E \| \ubar{\wt{\bm{w}}}_{k,i} \|^4 & \le [\max_k (1 - h_{k,1})] \cdot \max_k \E \| \ubar{\wt{\bm{w}}}_{k,i-1} \|^4  + \max_k [h_{k,2} \cdot 2(b  + \epsilon) \cdot \nu + h_{k,3}]
\end{align}
Let 
\begin{align}
\label{eqn:gamma4def}
\gamma_4 & \defeq \max_k (1 - h_{k,1}) = 1 - \min_k h_{k,1} \\
\label{eqn:theta4def}
\theta_4 & \defeq \max_k [h_{k,2} \cdot 2(b  + \epsilon) \cdot \nu + h_{k,3}]
\end{align}
where $b$ is from \eqref{eqn:mumaxdef}. We can then use  \eqref{eqn:bound4thorderequmax1} to write for large enough $i$ that
\begin{align}
\label{eqn:bound4thorderequmax2}
\max_k \E \| \ubar{\wt{\bm{w}}}_{k,i} \|^4 & \le \gamma_4 \cdot \max_k \E \| \ubar{\wt{\bm{w}}}_{k,i-1} \|^4 + \theta_4
\end{align}
Therefore, the fourth-order moment of the individual error is governed by \eqref{eqn:bound4thorderequmax2}. Whenever $|\gamma_4| < 1$, the quantity $\max_k \E \| \ubar{\wt{\bm{w}}}_{k,i} \|^4$ will have a bounded value asymptotically. In order to guarantee $|\gamma_4| < 1$, it is sufficient to have
\begin{align}
\label{eqn:stability4thordercond}
0 & < 4\bar{\mu}_k^{(1)} \lambda_{k, \min} - 2 \bar{\mu}_k^{(2)} (2\lambda_{k, \min}^2 + \lambda_{k, \max}^2 + 4\alpha)   - \bar{\mu}_k^{(4)} (\lambda_{k, \max}^4 + 8\alpha \lambda_{k, \max}^2 + 3 \alpha^2) < 2
\end{align}
for all $k$. This condition can be guaranteed by the sufficient conditions:
\begin{subequations}
\begin{align}
\label{eqn:stability4thorder1}
4\bar{\mu}_k^{(1)} \lambda_{k, \min} & < 2 \\
\label{eqn:stability4thorder2}
\bar{\mu}_k^{(2)} (2\lambda_{k, \min}^2 + \lambda_{k, \max}^2 + 4\alpha) & < \bar{\mu}_k^{(1)} \lambda_{k, \min} \\
\label{eqn:stability4thorder3}
\bar{\mu}_k^{(4)} (\lambda_{k, \max}^4 + 8\alpha \lambda_{k, \max}^2 + 3 \alpha^2) & < \bar{\mu}_k^{(2)} (2\lambda_{k, \min}^2 + \lambda_{k, \max}^2 + 4\alpha)
\end{align}
\end{subequations}
Condition \eqref{eqn:stability4thorder1} is equivalent to
\be
\label{eqn:stability4thorder1new}
\bar{\mu}_k^{(1)} < \frac{1}{2 \lambda_{k, \min}}
\ee
Condition \eqref{eqn:stability4thorder2} holds if
\be
\label{eqn:stability4thorder2new}
\frac{\bar{\mu}_k^{(2)}}{\bar{\mu}_k^{(1)}} < \frac{\lambda_{k, \min}}{3\lambda_{k, \max}^2 + 4\alpha}
\ee
Condition \eqref{eqn:stability4thorder3} holds if
\be
\label{eqn:stability4thorder3new}
\frac{\bar{\mu}_k^{(4)}}{\bar{\mu}_k^{(2)}} < \frac{1}{\lambda_{k, \max}^2 + 4\alpha}
\ee
because
\be
\frac{\lambda_{k, \max}^2 + 4\alpha}{(\lambda_{k, \max}^2 + 4\alpha)^2} < \frac{2\lambda_{k, \min}^2 + \lambda_{k, \max}^2 + 4\alpha}{\lambda_{k, \max}^4 + 8\alpha \lambda_{k, \max}^2 + 3 \alpha^2}
\ee
Since, for any random variable $\bm{\mu}_k(i)$,
\be
\label{eqn:boundmumoments0}
[\bar{\mu}_k^{(1)}]^2 \le \bar{\mu}_k^{(2)}, \qquad
[\bar{\mu}_k^{(2)}]^2 \le \bar{\mu}_k^{(4)}
\ee
it is straightforward that 
\be
\label{eqn:boundmumoments}
\max \left\{ [\bar{\mu}_k^{(1)}]^2, \left( \frac{\bar{\mu}_k^{(2)}}{\bar{\mu}_k^{(1)}} \right)^2, \frac{\bar{\mu}_k^{(4)}}{\bar{\mu}_k^{(2)}} \right\} \le \frac{\bar{\mu}_k^{(4)}}{[\bar{\mu}_k^{(1)}]^2}
\ee
On the other hand, it can be verified that
\be
\frac{ \lambda_{k, \min}^2 }{( 3\lambda_{k, \max}^2 + 4\alpha)^2 } < 
\min \left\{ \frac{1}{4 \lambda_{k, \min}^2}, \frac{1}{\lambda_{k, \max}^2 + 4\alpha} \right\} 
\ee
Therefore, if condition \eqref{eqn:boundstepsize4thorder} holds for all $k$, then \eqref{eqn:stability4thorder1new}--\eqref{eqn:stability4thorder3new} hold, and $|\gamma_4| < 1$ holds. Using \eqref{eqn:boundmumoments} and the new definition of $\nu$ in \eqref{eqn:nunewdef}, we obtain
\be
\label{eqn:boundmumoments2}
\bar{\mu}_k^{(1)} \le \nu, \quad \bar{\mu}_k^{(2)} \le \nu^2, \quad \bar{\mu}_k^{(4)} \le \nu^4
\ee
Using \eqref{eqn:boundmumoments2}, we have
\be
\label{eqn:hk2and3upperbound}
h_{k,2} \le 16 \sigma_v^2 \nu^2 (1 + \lambda_{k, \max}^2 \nu^2 ), \quad h_{k,3} \le 12 \sigma_v^4 \nu^4
\ee
It is worth noting that the new definition of $\nu$ in \eqref{eqn:nunewdef} bounds the old definition in \eqref{eqn:mumaxdef} from above since
\be
\label{eqn:oldnulessnewnu}
\frac{\bar{\mu}_k^2 + c_{\mu,k,k}}{\bar{\mu}_k} = \frac{\bar{\mu}_k^{(2)}}{\bar{\mu}_k^{(1)}} \le \frac{ \sqrt{\bar{\mu}_k^{(4)}} }{ \bar{\mu}_k^{(1)} } 
\ee
due to \eqref{eqn:boundmumoments}. It is easy to verify that 
\be
\label{eqn:newboundlessoldbound}
\frac{ \lambda_{k, \min} }{3\lambda_{k, \max}^2 + 4\alpha} < \frac{\lambda_{k,\min}}{\alpha + \lambda_{k,\max}^2}
\ee
With \eqref{eqn:oldnulessnewnu} and \eqref{eqn:newboundlessoldbound}, it is obvious that  \eqref{eqn:boundstepsize4thorder} implies \eqref{eqn:meansquarestabilitycond3}.

When $|\gamma_4| < 1$, the recursive inequality \eqref{eqn:bound4thorderequmax2} leads to
\be
\label{eqn:bound4thordermoment}
\limsup_{i\rightarrow\infty} \left[ \max_k \E \| \ubar{\wt{\bm{w}}}_{k,i} \|^4 \right] \le \frac{\theta_4}{1 - \gamma_4}
\ee
Substituting \eqref{eqn:hk2def} and \eqref{eqn:hk3def} into \eqref{eqn:theta4def} yields
\begin{align}
\label{eqn:boundtheta4}
\theta_4 & \le \max_k [ 16 \sigma_v^2 (\bar{\mu}_k^{(2)} + \bar{\mu}_k^{(4)} \lambda_{k, \max}^2) \cdot 2(b  + \epsilon) \nu + 12 \sigma_v^4 \bar{\mu}_k^{(4)} ] \nn \\
& = \max_k \left[ 32 \sigma_v^2 \bar{\mu}_k^{(2)} \left(1  +  \frac{ \bar{\mu}_k^{(4)} }{\bar{\mu}_k^{(2)}} \lambda_{k, \max}^2 \right) (b  +  \epsilon) \nu  +  12 \sigma_v^4 \bar{\mu}_k^{(4)} \right]
\end{align}
where $\nu$ is given by \eqref{eqn:nunewdef}. Using \eqref{eqn:boundstepsize4thorder} and \eqref{eqn:boundmumoments}, we have
\be
\label{eqn:boundmu4mu2lambda2}
\frac{ \bar{\mu}_k^{(4)} }{\bar{\mu}_k^{(2)}} \lambda_{k, \max}^2 < \frac{ \lambda_{k, \max}^2 \lambda_{k, \min}^2 }{( 3\lambda_{k, \max}^2 + 4\alpha)^2 } \le \frac{ \lambda_{k, \max}^4 }{( 3\lambda_{k, \max}^2)^2 } = \frac{1}{9}
\ee
Substituting \eqref{eqn:boundmu4mu2lambda2} into \eqref{eqn:boundtheta4} yields
\begin{align}
\label{eqn:boundtheta4_1}
\theta_4 & \le \max_k \left[ 32 \sigma_v^2 \bar{\mu}_k^{(2)} \frac{10}{9} (b + \epsilon) \nu + 12 \sigma_v^4 \bar{\mu}_k^{(4)} \right] \nn \\
& \stackrel{(a)}{\le} \max_k \left[ 12 \sigma_v^2 \bar{\mu}_k^{(2)} \left( 3 b \nu + \sigma_v^2 \frac{ \bar{\mu}_k^{(4)} }{\bar{\mu}_k^{(2)}} \right) \right] \nn \\
& \stackrel{(b)}{\le} \max_k \left[ 12 \sigma_v^2 \bar{\mu}_k^{(2)} ( 3 b \nu + \sigma_v^2 \nu^2 ) \right] \nn \\
& = \max_k \left[ 12 \sigma_v^2 \bar{\mu}_k^{(1)} \frac{ \bar{\mu}_k^{(2)} }{\bar{\mu}_k^{(1)}} ( 3 b \nu + \sigma_v^2 \nu^2 ) \right] \nn \\
& \stackrel{(c)}{\le} \max_k \left[ 12 \sigma_v^2 \bar{\mu}_k^{(1)} \nu^2 ( 3 b + \sigma_v^2 \nu ) \right]
\end{align}
where step (a) is by choosing $\epsilon \le b/80$; and steps (b) and (c) are by using \eqref{eqn:nunewdef} and \eqref{eqn:boundmumoments}. Substituting \eqref{eqn:stability4thorder2} and \eqref{eqn:stability4thorder3} into \eqref{eqn:hk1def} yields
\be
\label{eqn:boundhk1}
h_{k,1} \ge \bar{\mu}_k^{(1)} \lambda_{k, \min}
\ee
It follows from \eqref{eqn:gamma4def} and \eqref{eqn:boundhk1} that 
\be
\label{eqn:boundhk1_1}
1 - \gamma_4 = \min_k h_{k,1} \ge \min_k [\bar{\mu}_k^{(1)} \lambda_{k, \min}] \ge \min_k \bar{\mu}_k^{(1)} \cdot \min_k \lambda_{k, \min}
\ee
Substituting \eqref{eqn:boundtheta4_1} and \eqref{eqn:boundhk1_1} into \eqref{eqn:bound4thordermoment}, we arrive at
\begin{align}
\label{eqn:bound4thordermoment1}
\limsup_{i\rightarrow\infty} \left[ \max_k \E \| \ubar{\wt{\bm{w}}}_{k,i} \|^4 \right] & \le \frac{ 12 \sigma_v^2 \nu^2 ( 3 b + \sigma_v^2 \nu ) \cdot \max_k \bar{\mu}_k^{(1)}}{\min_k \bar{\mu}_k^{(1)} \cdot \min_k \lambda_{k, \min}} \nn \\
& \le \frac{ 12 \sigma_v^2 ( 3 b + \sigma_v^2 \nu )}{\min_k \lambda_{k, \min}} \frac{\max_k \bar{\mu}_k^{(1)}}{\min_k \bar{\mu}_k^{(1)}} \nu^2 \nn \\
& \le \frac{ 12 \kappa \sigma_v^2 ( 3 b + \sigma_v^2 \nu )}{\min_k \lambda_{k, \min}} \nu^2
\end{align}
where we used \eqref{eqn:kappadef} in the last step. From  \eqref{eqn:boundstepsize4thorder} and \eqref{eqn:nunewdef}, it is easy to verify that
\be
\label{eqn:boundnu}
\nu < \max_k \frac{ \lambda_{k, \min} }{ 3\lambda_{k, \max}^2 + 4\alpha } \le \frac{1}{3  \min_k \lambda_{k, \min}}
\ee
Then, from \eqref{eqn:mumaxdef} and \eqref{eqn:boundnu}, we obtain
\be
\label{eqn:bound6bsigmavnu}
3 b + \sigma_v^2 \nu \le \frac{3 \kappa \sigma_v^2 }{\min_k \lambda_{k, \min}} + \frac{\sigma_v^2}{3  \min_k \lambda_{k, \min}} < \frac{3 \sigma_v^2 ( \kappa + 1)}{\min_k \lambda_{k, \min}}
\ee
Therefore, we obtain from \eqref{eqn:bound4thordermoment1} and \eqref{eqn:bound6bsigmavnu} that
\begin{align}
\label{eqn:bound4thordermoment2}
\limsup_{i\rightarrow\infty} \left[ \max_k \E \| \wt{\bm{w}}_{k,i} \|^4 \right] \le b_4^2 \cdot \nu^2 = O(\nu^2)
\end{align}
due to the identity $\| \ubar{w} \|^4 = 4 \cdot \| w \|^4$, where $b_4$ is given by \eqref{eqn:nunewdef}.

\end{document}